\documentclass[9pt,twocolumn,twoside]{osajnl}

\usepackage{gnuplot-lua-tikz}
\usepackage[caption=false,font=footnotesize]{subfig}
\usepackage{adjustbox}
\usepackage{graphicx}
\usepackage{enumitem}
\usepackage{textcomp}
\usepackage{xcolor}
\usepackage{mathtools}
\usepackage[symbol]{footmisc}
\usepackage{array, boldline, makecell, booktabs}
\usepackage{amsmath,amssymb,amsfonts}
\usepackage{amsthm}
\usepackage{algorithm}
\usepackage[algo2e,ruled,vlined,linesnumbered]{algorithm2e} 
\usepackage{array, boldline, makecell, booktabs}
\usepackage[symbol]{footmisc}
\usepackage{mathtools}
\usepackage{comment}
\usepackage{flushend}

\usepackage{booktabs}
\usepackage{multirow}

\journal{jocn}

\title{Experimental Evaluation of an SDN Controller for\\ Open Optical-circuit-switched Networks}

\author[1,3,*]{Kazuya Anazawa}
\author[1,**]{Takeru Inoue}
\author[1]{Toru Mano}
\author[2]{Hiroshi Ou}
\author[2]{Hirotaka Ujikawa}
\author[4]{Dmitrii Briantcev}
\author[4]{Sumaiya Binte Ali}
\author[4]{Devika Dass}
\author[1]{Hideki Nishizawa}
\author[1]{Yoshiaki Sone}
\author[5]{Eoin Kenny}
\author[4]{Marco Ruffini}
\author[4]{Daniel Kilper}
\author[3]{Eiji Oki}
\author[1]{Koichi Takasugi}

\affil[1]{NTT Network Innovation Laboratories, 1-1 Hikarinooka, Yokosuka-shi, Kanagawa-ken, 239-0847, Japan}
\affil[2]{NTT Access Service System Laboratories, 1-1 Hikarinooka, Yokosuka-shi, Kanagawa-ken, 239-0847, Japan}
\affil[3]{Kyoto University, Yoshida-honmachi, Sakyo-ku, Kyoto, 606-8501, Japan}
\affil[4]{TRINITY College Dublin, Ireland}
\affil[5]{HEAnet, Dublin, Ireland}
\affil[*]{kazuya.anazawa@ntt.com}
\affil[**]{He is currently an associate professor at University of Yamanashi, 4-4-37 Takeda, Kofu-shi, Yamanashi-ken, 400-8510, Japan}

\begin{abstract}
	Open optical networks have been considered to be important for cost-effectively building and operating the networks. Recently, the optical-circuit-switches (OCSes) have attracted industry and academia because of their cost efficiency and higher capacity than traditional electrical packet switches (EPSes) and reconfigurable optical add drop multiplexers (ROADMs). Though the open interfaces and control planes for traditional ROADMs and transponders have been defined by several standard-defining organizations (SDOs), those of OCSes have not. Considering that several OCSes have already been installed in production datacenter networks (DCNs) and several OCS products are on the market, bringing the openness and interoperability into the OCS-based networks has become important. Motivated by this fact, this paper investigates a software-defined networking (SDN) controller for open optical-circuit-switched networks. To this end, we identified the use cases of OCSes and derived the controller requirements for supporting them. We then proposed a multi-vendor (MV) OCS controller framework that satisfies the derived requirements; it was designed to quickly and consistently operate fiber paths upon receiving the operation requests. We validated our controller by implementing it and evaluating its performance on actual MV-OCS networks. It satisfied all the requirements, and fiber paths could be configured within $1.0$ second by using our controller.
\end{abstract}

\setboolean{displaycopyright}{false} 

\begin{document}
	\maketitle
	\section{Introduction}
	Bringing the openness and interoperability into the optical networks has been considered as key for improving network operations~\cite{le2022operator}. Several standard-defining organizations (SDOs) have been working on the definitions of open interfaces and a software-defined networking (SDN) controller towards open optical networks~\cite{vilalta2021experimental, openroadm, openconfig, tpce}. For example, the data models for traditional reconfigurable optical add drop multiplexers (ROADMs) and transponders have been standardized in OpenConfig~\cite{openconfig} and OpenROADM Multi-source Agreement (MSA)~\cite{openroadm}.
	
	Recently, optical-circuit-switches (OCSes), which perform optical switching at a fiber layer, have been considered for installation on various networks such as intra-datacenter networks (DCNs)~\cite{sato2023optical, poutievski2022jupiter, dukic2020beyond, wang2024leaf, liu2023lightwave, wang2023topoopt}, wide-area networks (WANs)~\cite{dukic2020beyond, matsuo2023architecture}, and metro-access converged optical networks~\cite{kani2025disaggregation, kaneko2024photonic, igf}. The benefits of deploying OCSes are wide-ranging. Their typical advantages are higher energy efficiency and transparency than traditional electrical packet switches (EPSes) because the OCSes require no optical-electrical-optical (OEO) conversion for signal switching and have no queuing delay due to packet processing. Another advantage is capital expenditure (CAPEX) reduction. For example, OCS-based WANs can improve transmission performance because OCSes have much lower insertion loss than the traditional ROADMs and incur no filter penalty~\cite{shiraki2021design, matsuo2023architecture}. Thus, the number of required amplifiers can be reduced, and Quality of Transmission (QoT) can be improved~\cite{dukic2020beyond}. Though OCSes have these advantages, the unified control interface for managing multi-vendor (MV) OCSes and their controller have not been investigated and standardized. For network operators, vendor-neutral OCS management is inevitable to reduce the development cost and operational expenditure (OPEX).
	
	In this paper, we investigated an architecture and implementation design of an SDN controller called \textit{MV-OCS controller} for supporting various OCS use cases such as OCS-based (i) intra-DCNs and (ii) WANs, and (iii) authentication and link-by-link probing using OCSes. To this end, we carefully analyzed the controller requirements for supporting these use cases and the services that should be offered by the controller. On the basis of them, we designed an MV-OCS controller architecture and its sophisticated implementation design including the north-bound interface (NBI), south-bound interface (SBI), and internal functions. Hereinafter, we collectively call the network devices attached to the OCS-based networks [e.g., Top-of-Rack (ToR) switches and transponders] \emph{terminals}. The contributions of this paper are summarized as follows.
	\begin{itemize}
		\item This paper, for the first time, investigates the MV-OCS controller framework for open optical-circuit-switched networks. To this end, we identified the OCS use cases and derived the controller requirements for supporting them.
		\item This paper proposes a framework for the MV-OCS controller (i.e., architecture and implementation design) that satisfies all the derived controller requirements. The NBI for managing OCS-based networks, the unified SBI for operating MV-OCSes, and the internal functions and mechanisms for controlling fiber paths safely, quickly, and automatically upon receiving operation requests were designed.
		\item We clarified the feasibility and validity of the MV-OCS controller’s architecture by implementing it and evaluating its performance on actual MV-OCS networks. It satisfied all the derived controller requirements, and fiber paths were controlled within $1.0$ second. This paper also demonstrates the MV-OCS controller’s applicability through experiments in a live production field environment.
	\end{itemize}
	Considering that the SDN controllers are often defined and implemented for each layer~\cite{giorgetti2023enabling, gifre2022experimental}, designing one for optical-circuit-switched networks (i.e., fiber layer) has great importance to effectively manage the entire networks.
	
	The rest of this paper is organized as follows. Section~\ref{sec:related-works} reviews related works. Section~\ref{sec:usecases} identifies the OCS use cases and the challenges for realizing the MV-OCS controller. Section~\ref{sec:req} identifies the MV-OCS controller requirements. Section~\ref{sec:design} proposes the MV-OCS controller architecture and its implementation design. Section~\ref{sec:experimental-validation} evaluates the functionality and performance of our MV-OCS controller on an actual testbed. Finally, Section~\ref{sec:concluding-remarks} concludes this paper and mentions future directions.
	
	\section{Related Works}
	\label{sec:related-works}
	Though extensive efforts have been made for open optical networking, no studies have detailed the requirements and framework for an MV-OCS controller as we review in this section.
	
	Giorgetti et al.~\cite{giorgetti2020control} studied the Open Networking Operating System (ONOS) controller for open and disaggregated optical line systems. The reference architecture of the controller as well as its performance and limitations were reported in detail. Similarly, Vilalta et al.~\cite{vilalta2021experimental} reported the SDN controller architecture for optical networks defined by multiple SDOs. Pederzolli et al.~\cite{pederzolli2017yamato} presented the SDN control plane for space-division multiplexing (SDM) optical networks. They extended the OpenDaylight SDN framework to manage optical connectivity services on SDM networks and validated them by emulation-based study. Borraccini et al.~\cite{borraccini2023experimental} studied a partially disaggregated optical network architecture and verified its control and data planes using open source software called GNPy. Though several efforts on SDN control and management for optical networks have been made, all of these works mainly focus on the ROADM-based optical networks, not OCS-based circuit-switched networks. In addition, the above works did not consider interoperability of MV-OCSes.
	
	Patronas et al.~\cite{patronas2025optical} explored the Layer-1 SDN control plane in high-performance computing (HPC) clusters. Though they mentioned the importance of interoperability as well as the ecosystem for MV-OCSes, the unified interfaces for MV-OCSes as well as the detailed architecture of the controller were not studied in detail. Takano et al.~\cite{takano2024fast} proposed a fast control plane for optical-circuit-switched systems. They demonstrated a micro-second level network configuration on large-scale networks utilizing EtherCAT. Ferguson et al.~\cite{ferguson2021orion} presented the detailed architecture of Google's SDN control plane. It operates Google's proprietary OCSes for their network operations~\cite{poutievski2022jupiter}. Lei et al.~\cite{lei2024open, lei2024lighthouse} presented an experimental control platform for operating diverse optical DCN architectures. However, these works do not consider the OCS-based networks consisting of commercial MV-OCSes and their interoperability; the unified SBI for managing MV-OCSes has not been studied. In addition, they only focused on the SDN controller for optical DCNs and did not cover the OCS use cases studied in this paper. We still lack the definitions of open interface for MV-OCSes and their controller framework.
	
	In our previous work~\cite{anazawa2024first}, we designed and implemented an SDN controller for OCS-based intra-DCNs. This paper extends it by extensively investigating the OCS use cases and detailing a MV-OCS controller framework to support them. In addition, the performance of our controller was experimentally evaluated on real OCS-based networks with MV-OCSes.
	
	\section{OCS use cases and challenges}
	\label{sec:usecases}
	This section first describes the OCS use cases. We then discuss technical challenges for supporting them by the MV-OCS controller.
	
	\subsection{Use cases}
	\label{subsec:ocs-based-networks}
	\subsubsection{Optical-circuit-switched intra-DCNs}
	The optical-circuit-switched intra-DCNs have been extensively studied due to higher energy efficiency and better transparency than traditional EPSes~\cite{sato2023optical, poutievski2022jupiter, liu2023lightwave}. While the current practice of the OCS DCN is single-layered (e.g., Google's Jupiter~\cite{poutievski2022jupiter}), the hierarchical OCS DCN has also been studied~\cite{taniguchi2024optical, oki2025design}. Thus, we overview both cases. Fig.~\ref{fig:usecases}(a) shows a single-layer OCS DCN (leftmost) and the hierarchical one (rightmost) that accommodate terminals such as ToR switches or servers. The fiber paths are often established or released to execute topology-engineering (ToE)~\cite{poutievski2022jupiter}. In Fig.~\ref{fig:usecases}(a), a fiber path between active terminals (yellow) can be established by configuring one or more OCSes. Once the established path fails, it should be restored immediately. In both OCS DCN architectures, alternative paths should be established for a redundant terminal as discussed in~\cite{patronas2025optical}. In the case of hierarchical OCS DCNs, alternative paths can also be established between original terminals by using other fibers if we can identify a link failure point (e.g., fibers between OCSes); the failure point might be located by existing schemes such as the one in~\cite{anazawa2025verification}. Note that the path failures must be reported from a terminal, not an OCS, since OCSes cannot monitor signal statuses, unlike EPSes. Though several OCSes with a power monitor function are on the market, they are prohibitive to use for intra-DCNs because such OCSes incur additional insertion loss while the link-loss budget of transceivers used in intra-DCNs is often very limited~\cite{patronas2025optical}.
	
	\subsubsection{Optical-circuit-switched WANs}
\begin{figure}
		\begin{minipage}{.5\linewidth}
			\centering
			\subfloat[OCS-based intra-DCNs.]
			{\label{main:a}\includegraphics[scale=.53]{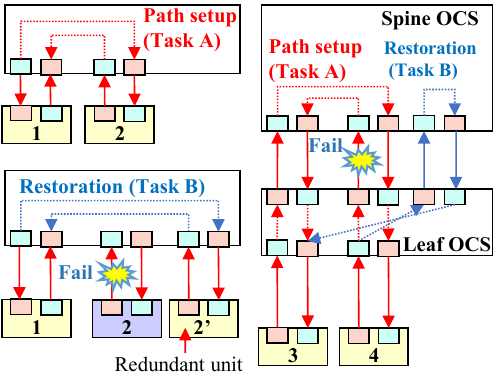}}
		\end{minipage}%
		\begin{minipage}{.5\linewidth}
			\centering
			\subfloat[OCS-based WANs.]
			{\label{main:b}\includegraphics[scale=.53]{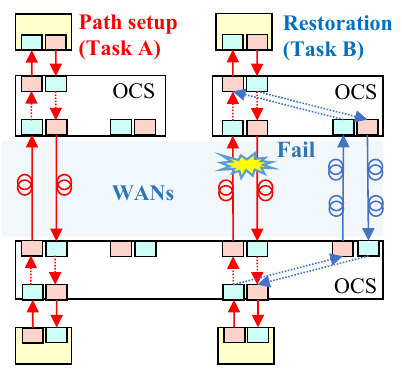}}
		\end{minipage}\par\medskip
		\centering
		\subfloat[Access-metro converged networks with OCSes.]
		{\label{main:c}\includegraphics[scale=.5]{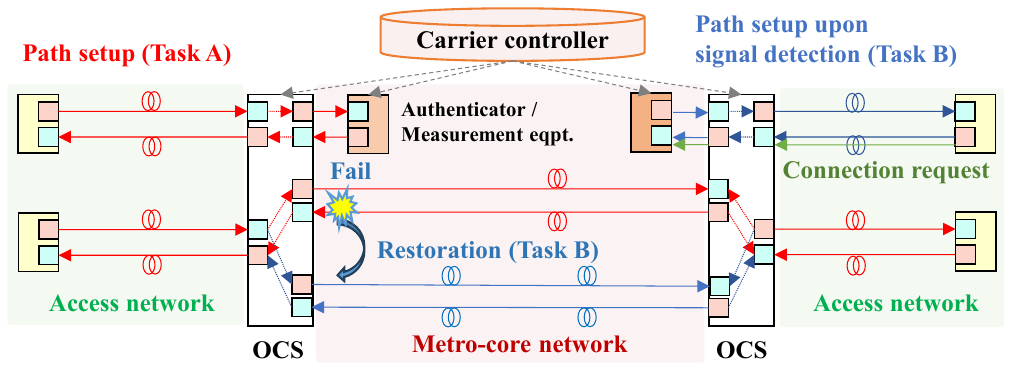}}
		\caption{The OCS use cases.}
		\label{fig:usecases}
	\end{figure}

	The optical-circuit-switched WANs have also been studied for reducing CAPEX and improving transmission performance~\cite{dukic2020beyond, yu2024network}. They interconnect buildings such as datacenters and telecommunication offices by OCSes instead of traditional ROADMs. As with the optical-circuit-switched intra-DCNs, the terminals attached to the networks are directly connected through fiber paths in optical-circuit-switched WANs, and they should be immediately restored upon path failures occurring as shown in Fig.~\ref{fig:usecases}~(b). However, unlike the intra-DCNs, signals from the terminals can traverse through many OCSes~\cite{dukic2020beyond}. This implies the necessity of configuring many OCSes for fiber-path control on this network. In addition, OCSes supporting a power monitor function can be deployed on line systems with amplifiers. In this case, the OCSes can monitor signal detection or degradation at a certain port and send the notification accordingly. Upon observing such events, fiber-path control (e.g., restoration) should be carried out automatically as with the traditional generalized multi-protocol label switching (GMPLS) networks~\cite{azodolmolky2011experimental}.
	
	\subsubsection{Authentication and link-by-link probing using OCSes}
	The OCSes can be used for carrying out authentication or link-by-link probing to accommodate the terminals outside the carrier networks~\cite{igf}. This use case, for example, could be applied in the metro-access converged network, which extends the wavelength-division-multiplexing (WDM) metro-core networks to access areas to provide large capacity lambda connections between user premises (e.g., transponders placed at datacenters)~\cite{kaneko2024photonic, igf, ruffini2014discus}. Fig.~\ref{fig:usecases}(c) shows the access-metro converged networks with OCSes. Recently, several working groups at the Innovative Optical and Wireless Network (IOWN) Global Forum have been studying a use case of flexibly providing the end-to-end lambda connections for users over access and metro-core networks. Here, not only lambda connections but also the underlying fiber paths are expected to be provided by utilizing OCSes, which are also called an APN Fiber Cross-connect (APN-FX)~\cite{igf, kani2025disaggregation}. The provisioning sequence of lambda connection is as follows~\cite{nishizawa2023dynamic}.
	\begin{enumerate}
		\item The user terminal sends signals to request a lambda connection to the carrier networks via access segments.
		\item The carrier controller detects the signal from a user terminal and switches it to the equipment (e.g., coherent transceivers used for bit-error-rate measurement) placed at the edge of metro-core networks by configuring OCSes. This process is required to authorize and estimate QoT of access segments~\cite{nishizawa2024fast}.
		\item The carrier controller provides a fiber path between user terminals by configuring OCSes placed at the carrier edge.
		\item The carrier controller finally provides a lambda connection between user terminals by tuning optical parameters (e.g., transmission mode, frequency, and output-power) for user terminals.
	\end{enumerate}
	
	The second and third processes can be automated by an MV-OCS controller; once the controller detects the user signal, the fiber path should be automatically controlled. Such automation is necessary for reducing OPEX.
	
	\begin{figure}[t]
		\centering
		\includegraphics[width=.45\textwidth]{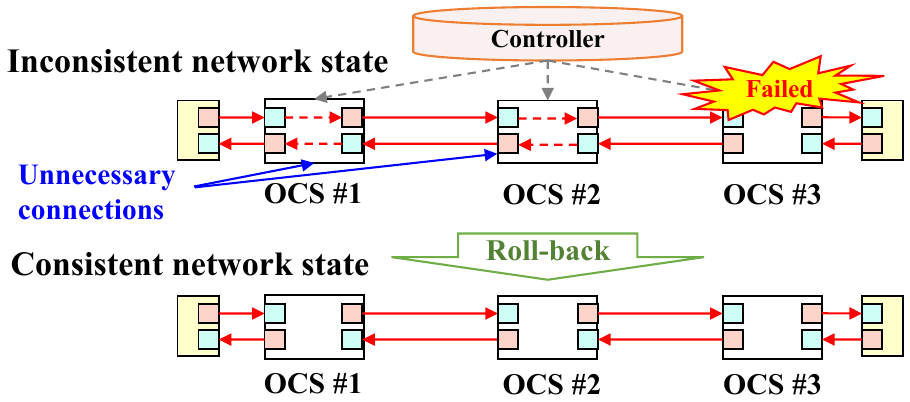}
		\caption{Inconsistent network state due to operation failure of OCS \#3. In this case, normal OCSes \#1 and \#2 should be rolled back to the previous network state (consistent network state).}
		\label{fig:issue}
	\end{figure}
	
	\subsection{Challenges on supporting OCS use cases}
	\label{subsec:task-analysis}
	This subsection details the challenges on supporting OCS use cases by analyzing operation tasks for them.
	
	\subsubsection{Fiber-path control}
	Fiber-path control is an essential task in each use case. Fiber paths between source and destination terminals should be established or released when a service creation or termination request from an operator is issued.
	This task is referred to as \textbf{Task A}. Here, we have to address three challenges for \textbf{Task A}.
	
	The first is how to define the NBI for fiber-path control on various OCS-based networks. As we reviewed in Section~\ref{sec:related-works}, the existing controller focused only on optical DCNs or ROADM networks. We thus need to define the NBI for this task.
	
	The second is how to configure MV-OCS in a unified manner. Though several device models (e.g., transponders or ROADMs) have been standardized, that of OCSes has not. This leads to vendor lock-in or increases the development cost of the controller due to the necessity of various kinds of OCS drivers. We thus need to design the unified SBI for operating MV-OCSes.
	
	The third is how to control fiber paths safely and fast. As shown in Fig.~\ref{fig:usecases}, we need to configure one or more OCSes when controlling a fiber path. In this case, if no OCSes can be successfully configured (e.g., due to equipment failure), the network state and the other normal OCSes should be rolled back before the path operation. Otherwise, unnecessary internal connections are left on several OCSes, which could lead to inconsistent network state and unexpected failures during operations. An example of the inconsistent network issue with three-node OCS-based networks is illustrated in Fig.~\ref{fig:issue}.
	Another issue is that controlling OCSes one by one takes time considering that OCS configurations via traditional network configuration protocols (NETCONF) take hundreds to thousands of milliseconds~\cite{suzuki2022automatic, vilalta2021experimental, pederzolli2017yamato}. This problem could be serious when many OCSes should be configured for fiber-path control like OCS-based WANs (Section~\ref{sec:usecases}.\ref{subsec:ocs-based-networks}). Thus, we need a sophisticated implementation design that satisfies both safe and fast fiber-path configuration.
	
	\subsubsection{Fiber-path setup upon signal detection or degradation}
	\label{subsubsec:signal-detection}
	Automatic fiber-path control upon signal detection or degradation will also be important. This operation task is referred to as \textbf{Task B}. Specifically, the fiber-path setup upon signal detection is important especially for metro-access converged networks; the operators should quickly detect the signals from user terminals and set up fiber paths for authorization and QoT estimation of access segments~\cite{kaneko2024photonic, nishizawa2024fast}.
	
	Automatic fiber-path control upon signal degradation could be useful for restoration; once established paths fail, they should be restored immediately. Recall that, for intra-DCNs, path failures must be reported from an endpoint, not an OCS, since OCSes in intra-DCNs often cannot monitor signal statuses, unlike EPSes. On the other hand, OCSes deployed on WANs or access-metro converged networks could have power monitor functions on their Tx/Rx ports, so they could monitor signal intensity and send notifications when they observe the power degradation events.
	
	Here, we face the same challenges as in the previous subsection. We need the NBI for executing this task as well as the unified SBI for \textit{monitoring} and \textit{notifying} the power detection or degradation events at OCSes. We also face another challenge for this operation. Since the number of terminals connected to the networks can be large, we need to asynchronously and concurrently control fiber paths when multiple events occur simultaneously. However, the MV-OCS controller framework for achieving this operation has not been clarified.
	
	\subsubsection{Summary}
	To summarize, the following items should be considered for realizing the MV-OCS controller.
	\begin{enumerate}
		\item How to define the NBI for executing operation tasks on OCS-based networks.
		\item How to define the unified SBI for managing MV-OCSes.
		\item How to realize safe and fast fiber-path control.
		\item How to realize automatic fiber-path control upon signal detection or degradation events.
	\end{enumerate}
	
	\section{Controller Requirements}
	\label{sec:req}
	This section describes the requirements on the MV-OCS controller on the basis of the use case analysis in Section~\ref{sec:usecases}. The requirements described in the following subsections (Subsections~\ref{sec:req}.\ref{subsec:req-a} to~\ref{sec:req}.\ref{subsec:req-d}) are also referred to as requirements~\ref{sec:req}.\ref{subsec:req-a} to~\ref{sec:req}.\ref{subsec:req-d} hereinafter.
	
	\subsection{North-bound interface for whole network management}
	\label{subsec:req-a}
	The fiber paths should be established or restored by specifying their identifiers (e.g., service names) and source and destination terminals (i.e., A and Z points) to be consistent with the traditional SDN controllers such as Transport Path Computation Engine (TPCE)~\cite{tpce}. For the fiber-path release, the identifier of target path should be specified. This requirement makes it easy to integrate or collaborate with the traditional higher layer's controllers.
	In addition, the NBI as well as a series of sanity check mechanisms should be defined and implemented for avoiding configuration mistakes by the users.
	
	\subsection{South-bound interface for managing MV-OCSes}
	\label{subsec:req-b}
	The fiber paths should be controlled even if MV-OCSes are installed. This requirement leads to avoiding vendor lock-in and reduces development cost. In this case, we need to define the unified model for OCSes. In addition, we would need an abstraction layer that reduces the source lines of code (SLOC) of the OCS driver inside the controller. Here, the overhead time introduced by the abstraction layer should be less than the half of the processing and configuration time by vendor-proprietary APIs. This requirement is necessary to satisfy requirement~\ref{sec:req}.\ref{subsec:req-c} described in the following subsection.
	
	\begin{figure}[t]
		\centering
		\includegraphics[width=.42\textwidth]{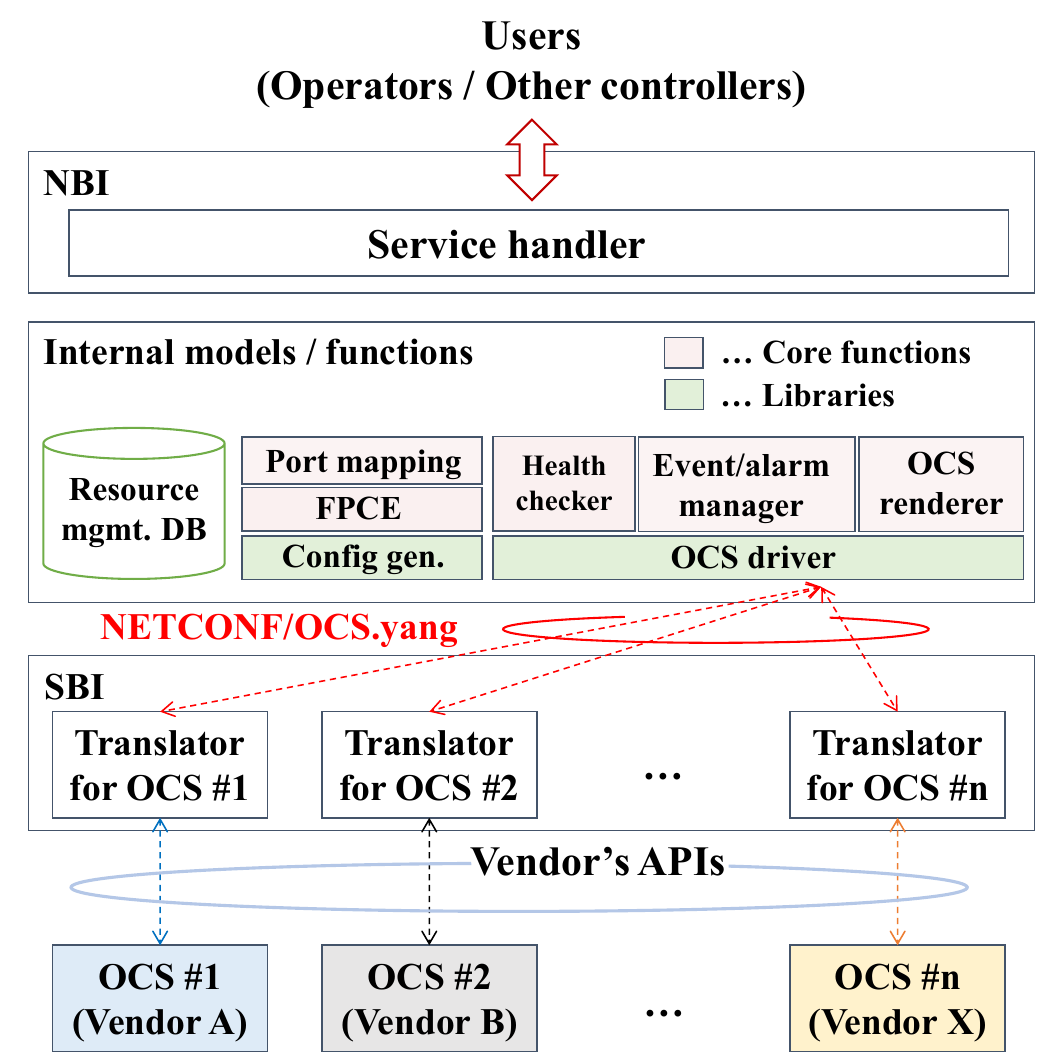}
		\caption{MV-OCS controller architecture.}
		\label{fig:controller-arch}
	\end{figure}
	\begin{figure}[t]
		\centering
		\includegraphics[width=.42\textwidth]{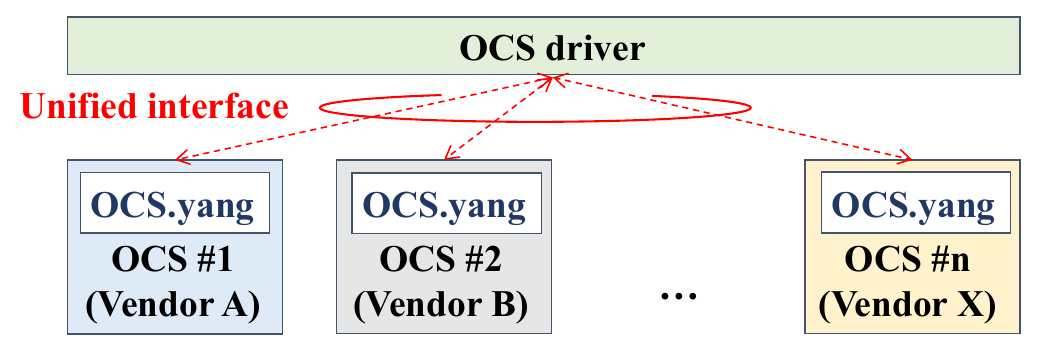}
		\caption{The goal of defining unified SBI. MV-OCSes should be operated by a unified interface.}
		\label{fig:sbi-goal}
	\end{figure}
	
	\subsection{Fast and safe fiber-path control}
	\label{subsec:req-c}
	The fiber paths should be controlled within minimal time for efficient operations. In addition, the network state should always be consistent to avoid unexpected failures. To this end, the OCSes should be atomically and concurrently controlled during fiber-path control. Since configuring OCSes via NETCONF takes hundreds to thousands of milliseconds~\cite{suzuki2022automatic, vilalta2021experimental, pederzolli2017yamato}, we target $1.0$ second for fiber path control.
	
	\subsection{Fiber-path control upon signal detection/degradation}
	\label{subsec:req-d}
	Once a signal is detected at a certain OCS port, the specified fiber path should be automatically established to accommodate terminals at user premises in metro-access converged networks. According to~\cite{nishizawa2024fast}, the lambda connection should be established in less than $10$ minutes. Thus, the fiber-path setup time should be sufficiently shorter than 10 minutes. As we mentioned, NETCONF operation could take hundreds to thousands milliseconds. Thus, we target $2.0$ seconds for fiber-path setup upon signal detection because at least two NETCONF operations are executed: handling NETCONF notification from OCSes and their configurations for fiber-path setup. As for the automatic restoration upon signal degradation, we target $3.0$ seconds because we need at least three NETCONF operations: handling NETCONF notification from OCSes, releasing the failed path, and establishing a new path.
	
	\section{MV-OCS controller framework}
	\label{sec:design}
	This section proposes an MV-OCS controller architecture and its implementation design. The overall architecture is shown in Fig~\ref{fig:controller-arch}. Since the traditional SDN controllers consist of three main function blocks (NBI, internal data models and functions, and SBI)~\cite{vilalta2021experimental}, we followed the same style. Note that the SBI translation layer was implemented inside our controller to operate MV-OCSes in a unified manner as shown in Fig.~\ref{fig:controller-arch}. In future, we will try to standardize the SBI so that it can be deployed on commercial MV-OCSes for their unified operation as shown in Fig.~\ref{fig:sbi-goal}. Hereinafter, we collectively call the client of our controller such as operators or other controller \textit{users}.
	
	\subsection{North-bound interface (NBI)}
	\label{subsec:nbi}
	We first give an overview of our defined NBI. It is classified into three categories: (1) network resource registration and management, (2) fiber-path control, and (3) event handling services, as listed in Table~\ref{table:nbi}. The service handler offers these services.
	
	\subsubsection{Network resource registration and management service}
	The users should register and manage information about OCSes, fiber links, terminals, and their physical topology via network resource registration and management services. For this operation, five interfaces were defined as listed in Table~\ref{table:nbi}. Since OCSes can only deflect signals from an Rx port to a Tx port, they cannot utilize a topology discovery protocol as EPSes~\cite{breitbart2004topology} or a link management protocol (LMP)~\cite{nadeau2005gmpls} as ROADMs. In addition, their availability status cannot be recognized by themselves unlike EPSes. Thus, the controller users need to provide the information about the fiber layer's resources and physical connection configurations of terminals (i.e., which terminals are attached to which ports of OCSes) with the controller. For an OCS registration, its identifier, connection information (e.g., host address and management port), and the sets of Tx/Rx ports should be given via \textbf{AddSwitch}(). The \textbf{AddTerminal}() can be used to register terminal; connection information should be provided so that the controller can await the notification from the terminals. For a link registration between OCSes or a pair of an OCS and a terminal, the identifier of source and destination objects as well as their port labels should be given via \textbf{AddLink}(). The \textbf{CreateNetwork}() interface allows users to register resource information all at once; it can be described in \emph{topology\_file} with JSON or YAML format. All of the registered information is stored and managed on the controller's database (DB).
	We also defined the \textbf{UpdateResourceStatus}() to update the status of each resource; each resource should be updated to \emph{unavailable status} if the users identify the network resource failures. The unavailable resources are ignored when calculating a fiber-path by the controller.
	
	\begin{table}
		\small
		\centering
		\caption{North-bound interfaces for whole network management}
		\begin{tabular}{l|l}
			\toprule
			NBI Category & Interfaces \\
			\midrule
			\multirow{5}{5pt}{Resource\\ registration}
			& \textbf{AddSwitch}(\textit{ocs\_id, conn\_info, tx\_ports, rx\_ports}) \\
			& \textbf{AddTerminal}(\textit{terminal\_id, conn\_info}) \\
			& \textbf{AddLink}(\textit{link\_id, src, dst, src\_port, dst\_port}) \\
			& \textbf{CreateNetwork}(\textit{topology\_file}) \\
			& \textbf{UpdateResourceStatus}(\textit{object\_id, object\_type}) \\
			\midrule
			\multirow{4}{5pt}{Path control}
			& \textbf{CreateFiberPath}(\textit{svc\_id, a, z [, pce\_alg, ocs\_list]}) \\
			& \textbf{DeleteFiberPath}(\textit{svc\_id}) \\
			& \textbf{RestoreFiberPath}(\textit{svc\_id, a, z [, pce\_alg, ocs\_list]}) \\
			& \textbf{UpdatePathAvailability}(\textit{svc\_id, status}) \\
			\midrule
			\multirow{5}{5pt}{Event/alarm\\handling}
			& \textbf{AddEvent}(\textit{event\_id, event\_type, ocs, port, threshold}) \\
			& \textbf{CreateAction}(\textit{act\_id, svc\_id, a, z [, pce\_alg, ocs\_list]}) \\
			& \textbf{DeleteAction}(\textit{act\_id, svc\_id}) \\
			& \textbf{CreateEventHandler}(\textit{event\_id, act\_id}) \\
			& \textbf{CreateAlarmHandler}(\textit{svc\_id, act\_id}) \\
			\bottomrule
		\end{tabular}
		\label{table:nbi}
	\end{table}

	\subsubsection{Fiber-path control service}
	\label{subsubsec:path-svc}
	The users can establish, release, or restore fiber paths by the fiber-path control service. The NBI for this service was designed to meet requirement~\ref{sec:req}.\ref{subsec:req-a}. Specifically, the users can establish fiber paths by specifying its service identifier, source and destination terminals via \textbf{CreateFiberPath}(). It also has the optional arguments, \emph{pce\_alg} and \emph{ocs\_list}. The users can specify a path computation algorithm in \emph{pce\_alg} (Dijkstra's shortest path is set by default). The \emph{ocs\_list} can be used to explicitly specify the list of OCSes on the path they desire to establish; the path establishment with the provided OCS list is prioritized when it is given. An established path can be released via \textbf{DeleteFiberPath}() by specifying its identifier. The \textbf{RestoreFiberPath}() is defined for restoring a fiber path between specified source and destination terminals. It first deletes the original path between specified terminals (\emph{a} and \emph{z}) on the basis of \textbf{DeleteFiberPath}() and then searches for and establishes a new path on the basis of \textbf{CreateFiberPath}(). In other words, \textbf{RestoreFiberPath}() internally wraps the \textbf{DeleteFiberPath}() and \textbf{CreateFiberPath}() to restore with one function call. These interfaces were defined as with service-create, service-delete, and service-restoration in a traditional TPCE controller~\cite{tpce}.
	
	To mitigate the failure caused by configuration mistakes of fiber paths, we also propose a basic sanity check mechanism and the necessary NBI for it on the controller.
	Specifically, after configuring OCSes for establishing or releasing fiber paths, the controller should check the operational state (i.e., actual internal connections) of OCSes by NETCONF GET operations~\cite{nc}. If the configuration set by users and state of an OCS do not match, the controller should raise \textbf{PathOperFailed} error (explained in Subsection~\ref{sec:design}.\ref{subsec:nbi}.\ref{subsubsec:exception}) and update the OCS status to \textbf{UNAVAILABLE}. In the case that the OCS behaves normally (i.e., no issues on its configuration and state), but it actually does not set up the path due to its silent or unknown failures, the controller could have no way of knowing this, but the controller users could do (e.g., due to loss of signal at the terminals). In this case, the users can recognize that the OCSes or links on the path could have failed, so their status should be temporarily \textbf{UNAVAILABLE}. To realize this operation, we defined the \textbf{UpdatePathAvailability}(\textit{svc\_id, status}) interface so that the controller users can register the availability of a path (whose id is \textit{svc\_id}). It also updates the status of all resources on the path to the specified \textit{status} (available or unavailable). Then, the users can try to establish another path.
	After the fiber-path configuration has successfully completed, the controller should start awaiting an asynchronous alarm (e.g., OpenConfig alarm) from the source and destination terminals as well as OCSes to immediately handle any failures on the path. The action toward mitigating the failure could be realized by the event and alarm handling services described in the following subsection.
	\begin{table}
		\small
		\centering
		\caption{List of exceptions that could be raised by the NBI}
		\begin{tabular}{l|l}
			\toprule
			Exceptions & Descriptions \\
			\midrule
			\textbf{AlreadyExist} & Same id of resource/path/event exists \\
			\textbf{ConnectionFailed} & Registered OCS could not be connected \\
			\textbf{NotFound} & Specified resource/path/event does not exit \\
			\textbf{InvalidRange} & Out of range value was given \\
			\textbf{BlockingOccured} & Path configuration failed due to blocking \\
			\textbf{PathOperFailed} & Path configuration failed due to OCS error\\
			\bottomrule
		\end{tabular}
		\label{table:error}
	\end{table}
	
	\subsubsection{Event and alarm handling service}
	Finally, the event and alarm handling services were defined to allow users to automatically carry out the desired fiber-path control when the user-defined events or alarm were observed. Five NBIs were defined for this service. The \textbf{AddEvent}() is used for registering user-defined events the users desire to observe. To this end, the users should provide \textit{event\_id}, \textit{event\_type} (i.e., signal detection or degradation), power threshold, and an OCS port the users desire to observe that event. The \textbf{CreateAction}() and \textbf{DeleteAction}() are used to register fiber-path control actions towards the events or alarms. The \textbf{CreateEventHandler}() creates event handlers to immediately execute actions specified by \textit{action\_id} when an event specified by \textit{event\_id} is observed. Similarly, \textbf{CreateAlarmHandler}() creates a handler to immediately execute actions specified by \textit{action\_id} when the controller observes the alarm from OCSes or terminals on a path whose service identifier is \emph{svc\_id}.
	\begin{figure}[t]
		\centering
		\includegraphics[width=.45\textwidth]{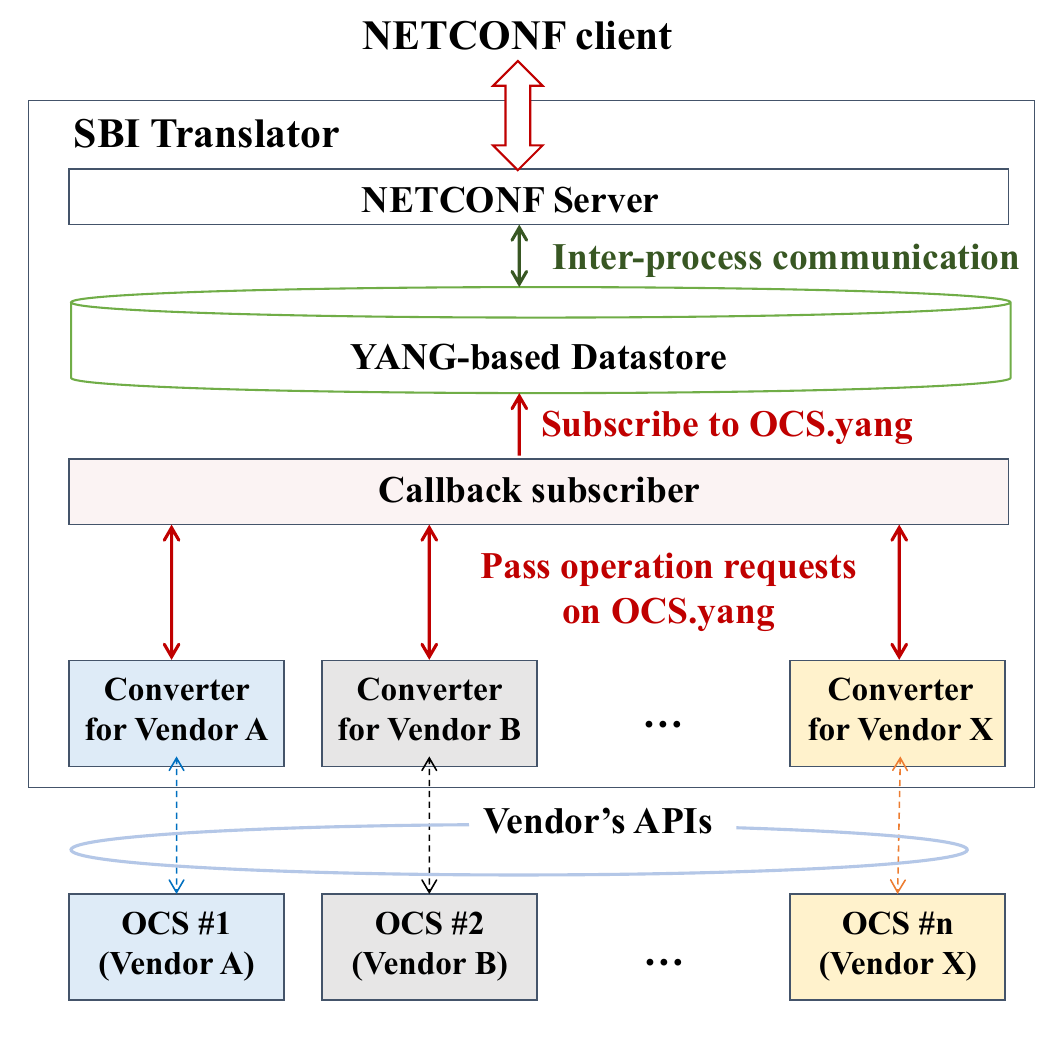}
		\caption{The function blocks of SBI translator.}
		\label{fig:sbi}
	\end{figure}
	
	\subsubsection{Exceptions raised by controller}
	\label{subsubsec:exception}
	The controller should raise exceptions when the users specified wrong or invalid argument(s) in each NBI or path operation failed. Thus, we defined a list of exceptions which could be raised by the controller (Table~\ref{table:error}). \textbf{AlreadyExist} error could be raised if the same identifier of a resource/path/event already exists when their registration; duplication of identifiers should be avoided to uniquely manage each object. \textbf{ConnectionFailed} error could be raised when the controller cannot access the OCS or terminal at the time of its registration. \textbf{NotFound} error could be raised when a specified resource/path/event has not been registered when their creation or deletion operations. \textbf{InvalidRange} error could be raised when the port or threshold is out of range. \textbf{BlockingOccured} error could be raised when a feasible path between specified two terminals does not exist at the time of path establishment or restoration due to the lack of network resources. \textbf{PathOperFailed} error could be raised when the path configuration or event handling cannot be successfully completed due to OCS failure.

	\subsection{South-bound interface (SBI)}
	To realize a unified SBI for managing MV-OCSes inside our controller (requirement~\ref{sec:req}.\ref{subsec:req-b}), we defined a Yet Another Next-Generation (YANG) data model called OCS.yang. We then designed an SBI translator that services OCS.yang over NETCONF.
	
	\subsubsection{YANG model for OCS management}
\begin{figure}[t]
		\centering
		\includegraphics[width=.5\textwidth]{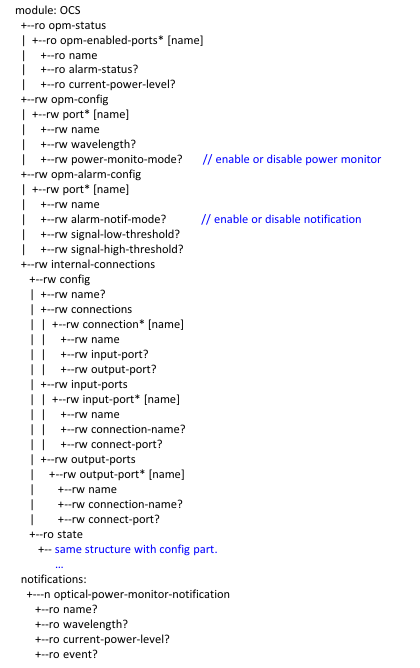}
		\caption{Tree view of our defined OCS.yang.}
		\label{fig:ocs-yang}
	\end{figure}

	Our defined YANG model is shown in Fig.~\ref{fig:ocs-yang}. It has a container and leaves for OCS configurations, state management, and notifications. For OCS configurations, leaves for configuring power monitor, power alarm, and internal connections leaves are defined. The internal connections on an OCS can be configured by specifying the connection name and Tx/Rx port pairs. When deleting an internal connection, the target connection name should be specified.
	Note that we could get \textit{actual} internal connections established on the OCSes by carrying out NETCONF GET operations for state leaf under internal-connections container.
	For power monitor configurations, the client can enable a power monitor for a certain port that they want to monitor the signal intensity. Furthermore, if the OCSes can monitor signal intensity with specific wavelengths, the client can also specify it. The current power level at a target port can be determined by the power-status leaf defined in the model. Regarding the alarm configurations, the client can configure OCSes so that they notify if a certain level of signal is detected or degraded. The threshold for such a signal detection and degradation should be given in signal-high-threshold and signal-low-threshold leaves.
	
	The proposed OCS.yang model manages internal connections in a unidirectional manner. Thus, the model is adoptable for OCSes that support unidirectional internal connection creation and deletion. To the best of our knowledge, the majority of commercial OCSes can support such a unidirectional internal connection creation and deletion (of course, the bidirectional internal connections can also be managed using the OCS.yang model). Note that the model does not rely on the switching mechanism of OCSes. In fact, we found that the robotic-arm, micro-electrical-mechanical-system, and piezo-actuator type commercial OCSes from four vendors could be managed by our model.
	
	\subsubsection{SBI translator}
	We then present an SBI translator that services OCS.yang over NETCONF. It configures MV-OCSes by translating an operation on OCS.yang into an operation by vendor-proprietary application programming interfaces (APIs). The architecture of the SBI translator is shown in Fig.~\ref{fig:sbi}. Each translator is dedicated for one OCS in the network. It consists of a NECTONF server, YANG-based datastore, callback subscriber, and converter. The OCS.yang is installed on the YANG-based datastore, and the NETCONF server is responsible for offering the operations on OCS.yang. The callback subscriber subscribes to the operations on OCS.yang; it immediately passes the operation requests on the OCS.yang to the converter, which converts the operations on OCS.yang into vendor proprietary API based operations for managing dedicated OCSes. The operation requests include the configurations of internal connection, power monitor, and power alarm, and a NETCONF GET operation to retrieve the state of OCSes.
	When integrating a new vendor's OCSes, we only need to implement the converter for those OCSes. All the other components are reusable due to their loosely-coupled architecture, so less effort is required for integrating new OCSes. Thanks to the translator, each OCS can be managed in a unified manner.
	
	\subsection{Implementation design}
	\label{subsec:proposed-design}
	This section describes the implementation design of internal functions for satisfying requirements~\ref{sec:req}.\ref{subsec:req-c} to~\ref{sec:req}.\ref{subsec:req-d}. The internal functions are composed of five main components: (1) Health checker, (2) Port mapping, (3) Fiber-Path Computation Engine (FPCE), (4) OCS renderer, and (5) Event and alarm manager.
	\begin{figure}[t]
		\centering
		\includegraphics[width=.48\textwidth]{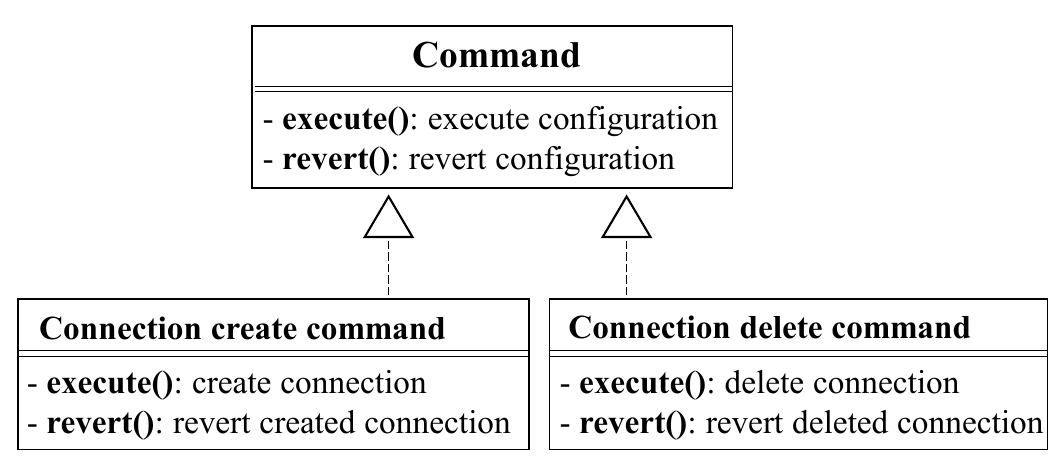}
		\caption{Command object for configuring OCSes.}
		\label{fig:command}
	\end{figure}
	
	\subsubsection{Implementation design for fiber-path control}
	\label{subsubsec:path-control-design}
	We first present the implementation design for controlling fiber paths on the basis of our defined NBI. To the best of our knowledge, the internal data models for representing network resources at the fiber layer (that is, OCS nodes, ports, fiber links, and the topology) have been not defined in the traditional standards so far. Thus, we newly defined them by following the model described in~\cite{vilalta2021experimental}. Specifically, the node (OCS) is represented as its identifier (name) and the set of Tx/Rx ports, the fiber link is represented as its identifier and the connection endpoints (i.e., the source and destination nodes and their Tx/Rx ports), and the topology is defined as the set of nodes and links. In addition, the fiber-layer resources should have an attribute for representing their availability. For example, if the users have identified a port failure, it should be treated as an unavailable port. Thus, we also define the status attribute for the OCS nodes, ports, and fiber links; it takes enum values of \textbf{AVAILABLE} or \textbf{UNAVAILABLE}. The attribute can be managed by the \textbf{UpdateResourceStatus}() interface.
	The status of OCSes can also be managed and checked by health checker inside the controller by periodically sending NETCONF hello~\cite{nc} or test packets.
	Once they are registered or updated, the resource-management DB stores that information in accordance with the model.

	We then defined a component called \textit{port mapping} similar with TPCE controller~\cite{tpce}, which is responsible for abstracting the network resources and register the information on the resource-management DB in accordance with the internal models. Relying on it, the FPCE computes the fiber path between specified source and destination terminals by considering the status of each resource (i.e., unavailable resources are ignored for path computation).
	Here, several path computation techniques (e.g., Dijkstra algorithm based shortest path computation) could be considered. Thus, the FPCE should be implemented as a selectable and pluggable module on the controller. In this way, the users can select a desired path computation algorithm when establishing a fiber-path between arbitrary source and destination terminals.
	On the basis of the path computation results, FPCE internally generates a set of OCSes on the candidate paths and their configurations necessary for establishing each candidate path relying on the port mapping and configuration generator. If no feasible path has been found by FPCE, it should raise the \emph{BlockingOccured} error and inform controller users of it.
	Note that the configuration generator should group multiple internal connection configurations of an OCS into one configuration payload (e.g., XML payload) to reduce communication times with each OCS. This leads to fast fiber-path control. Finally, generated configurations are passed to the OCS renderer, which is responsible for concurrently and atomically configuring OCSes (this is explained in the following subsection). The information about the established path (i.e., path name, source and destination terminals, and OCS configurations for the path) is stored on the persistent volume in the controller's DB.
	This makes failure recovery process of the controller as well as fiber-path release and restore process easy.
	
	On the basis of the above internal data models and components, we can establish a fiber path by specifying source and destination terminals. In addition, the controller could perform a basic sanity check. Thus, the requirement~\ref{sec:req}.\ref{subsec:req-a} is satisfied.
	
	\begin{figure*}[t]
		\centering
		\includegraphics[width=1.0\textwidth]{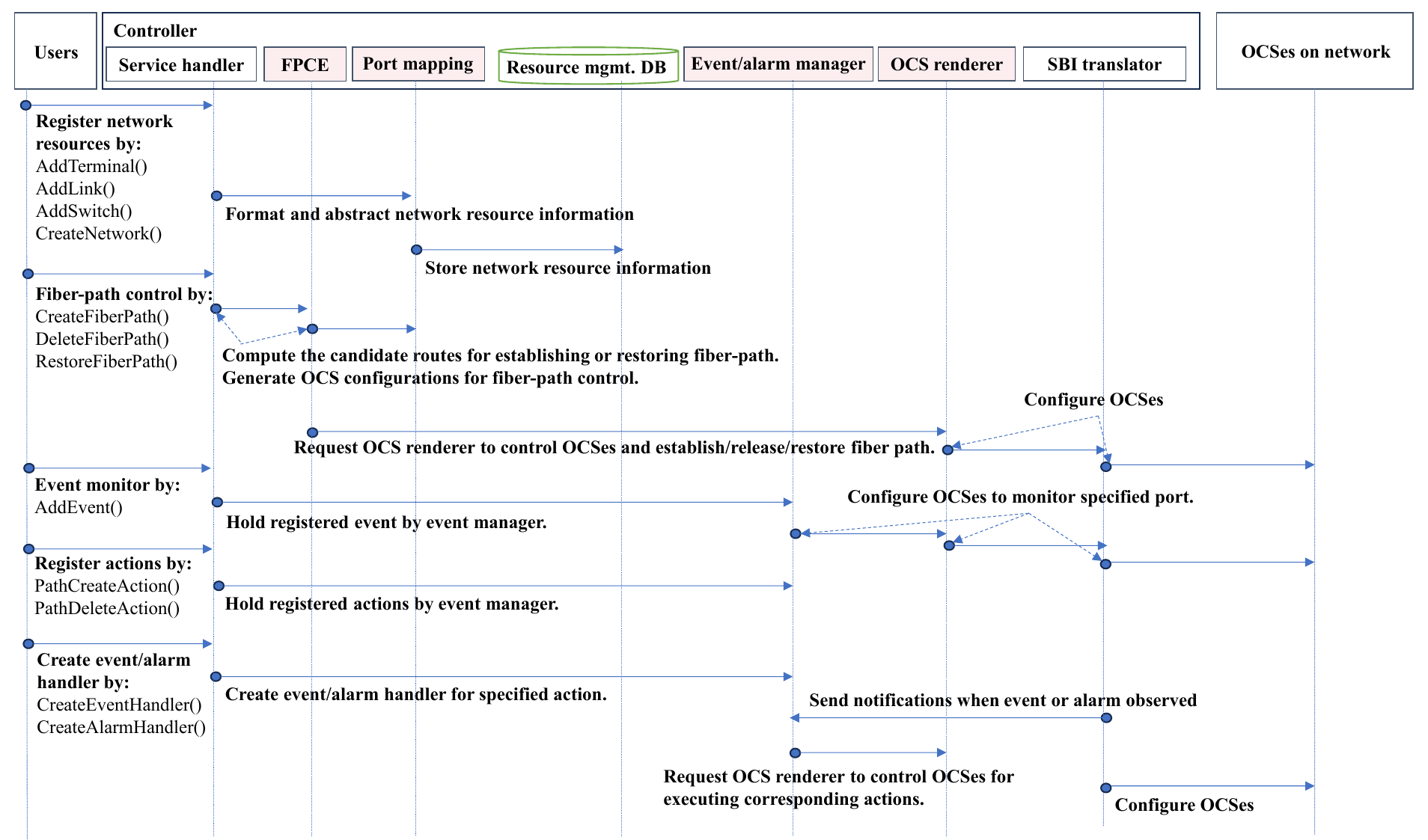}
		\caption{The operation flow by our controller.}
		\label{fig:oper-flow}
	\end{figure*}
	
	\subsubsection{OCS renderer for fast and safe fiber-path control}
	We defined a component called \textit{OCS renderer}, which atomically and concurrently configures multiple OCSes for fiber-path control to satisfy requirement~\ref{sec:req}.\ref{subsec:req-c}. It relies on the OCS driver inside the controller to control OCSes on the basis of our defined SBI. To atomically control multiple OCSes and support roll-back operations in the case of configuration failures, we applied a command pattern~\cite{command} for implementing OCS renderer. In this approach, the configuration to each OCS is wrapped as a command object, which consists of a configuration execution function (\textbf{execute}()) and a revert function (\textbf{revert}()) as shown in Fig.~\ref{fig:command}. This approach allows us to flexibly manage and execute the OCS configurations. Specifically, to atomically configure OCSes for fiber-path control, the OCS renderer generates an atomic command by combining multiple commands. Then, it concurrently calls the \textbf{execute}() function in each command to configure multiple OCSes for fiber-path control. If any command fails to execute, the OCS renderer immediately executes the \textbf{revert}() function to revert all the other commands. Here, the OCS renderer can know the operation failure by receiving the exceptions from our SBI (e.g., \textbf{RPCError} or \textbf{TimeoutError}). It also could know the configuration failure of an OCS on the basis of a sanity check mechanism in Subsection~\ref{sec:design}.\ref{subsec:nbi}.\ref{subsubsec:path-svc}. At the time OCS failures, the OCS renderer raises the \text{PathOperFailer} error.
	In this way, we can atomically configure OCSes for fiber-path control and satisfy requirement~\ref{sec:req}.\ref{subsec:req-c}.

	\subsubsection{Event and alarm manager for automatic fiber-path control}
	We finally defined a component called \textit{event and alarm manager}, which is responsible for handling the user-defined events and alarms to satisfy requirement~\ref{sec:req}.\ref{subsec:req-d}. It awaits the events and alarms from the network devices, and manages their handlers to automatically execute user-specified fiber-path control. To achieve this operation, we applied an event-driven approach. Specifically, once the user-defined events are registered, the event and alarm manager configures an OCS so that it notifies (publishes) when the events occurred. Then, the event and alarm manager awaits the notification from the OCS or terminals. Once it receives the notification, associated fiber-path control actions are immediately executed. By this event-driven approach, the fiber path can be immediately controlled after events have occurred. In addition, it ensures the scalability in terms of the number of events that should be handled because the event and alarm manager only awaits the event notification; it requires no periodical monitoring (polling process) of the OCSes and terminals. Furthermore, multiple handlers can be asynchronously invoked upon the events and are immediately executed.

	\subsection{Operation flow}
	The operation flow using our MV-OCS controller is depicted in Fig~\ref{fig:oper-flow}. First, the users need to register network resources and topology via network resource registration service. Once the service handler receives resource registration requests, the port mapping formats and abstracts the received resource information in accordance with the internal data models. Then, they are stored on the resource-management DB. After that, the users can establish, release, or restore fiber paths via a fiber-path control service. The FPCE computes the shortest fiber path between the specified source and destination terminals in accordance with the resource availability relying on the port mapping. It also generates configurations to OCSes for controlling fiber paths, and OCSes are configured via the OCS renderer and SBI translator. The users can also register the events and associated fiber-path control they desire to execute when the events occur. For this operation, the users can register events via \textbf{AddEvent}(). For example, the users can use this interface like \textbf{AddEvent}(\textit{Event\_A, signal\_detection, OCS\_\#1, Rx\_\#1, $-1.0$-dBm}), which can be translated to \textit{the Rx port \#1 of OCS \#1 observes the signal higher than $-1.0$ dBm}. Once the service handler receives this request, the event and alarm manager holds it and operates OCSes to monitor the event. Then, the users can register fiber-path control actions they desire to execute once the events are observed. The example is \textbf{CreateAction}(\textit{Act\_A, Service\_A, A, Z}), which can be translated to \textit{establish a fiber path between terminals A and Z with the service name of Service\_A}. Finally, the users explicitly register the event-action pairs by \textbf{CreateEventHandler}(). Once the users execute \textbf{CreateEventHandler}(\textit{Event\_A, Act\_A}), our controller creates an event handler that performs Act\_A as soon as the Event\_A is observed.
	
	\subsection{Recovery mechanism of controller from its failure}
	We finally describe how the controller recovers from its failure at runtime and resumes its services. Suppose the controller process is managed by a service manager such as Systemd or Kubernetes. When the controller fails, the service manager should restart or launch a new controller process. When the controller is launched, it first retrieves the registered resource information as well as the fiber-path and OCS configurations stored on the DB; they should always be retrieved because they are stored on the persistent volume as explained in Section~\ref{sec:design}.\ref{subsec:proposed-design}.\ref{subsubsec:path-control-design}. Then, the controller checks if actual OCS configurations match them; OCSes with mismatched configurations should be re-configured or marked as \textbf{UNAVAILABLE} due to the mismatches (it depends on the operation policy). After this reconciliation process, the controller resumes the services.
	
	\section{Experimental Validation}
	\label{sec:experimental-validation}
	This section implements the MV-OCS controller and evaluates it on the testbed with actual MV-OCSes and large-scale emulated networks.
	Hereinafter, we refer to our controller as \textit{L0- (Layer-0) controller}.
	
	\subsection{Implementation}
	We implemented our MV-OCS controller on the basis of the architecture presented in Section~\ref{sec:design}. The Python language (v3.11) and its libraries were used for the implementation.
	
	The NBI and internal functions were implemented as follows. To implement the service handler, we adopted a Google Remote Procedure Call (gRPC) as the communication protocol due to its high-performance RPC~\cite{vilalta2020uabno, vilalta2021experimental}. For the resource-management DB, we utilized Redis DB~\cite{redis}. The FPCE was realized by using NetworkX library~\cite{nx}. We implemented the OCS renderer by utilizing asyncio\_gather~\cite{coroutines} and run\_in\_executor~\cite{event-loop} APIs to concurrently execute multiple OCS commands. Since the NBI was implemented in accordance with Table~\ref{table:nbi}, the first requirement~\ref{sec:req}.\ref{subsec:req-a} was satisfied.
	
	The SBI translator for each OCS was implemented as a container application and deployed on lightweight Kubernetes called k3s~\cite{k3s}. Though each translator was deployed as one of the components in our controller, it also can be deployed on actual OCSes easily due to the high portability nature of container applications. This implementation allows OCS vendors to try our SBI translator. Each function block of the SBI translator was implemented as follows. For the YANG-based datastore, we utilized an open source software called Sysrepo~\cite{sysrepo, sysrepo-python}. For the NETCONF server, we used Sysrepo's brother project called netopeer2~\cite{np2}, which uses Sysrepo as the backend. As for the NETCONF and SSH client for communicating with OCSes via vendor-proprietary APIs, we used libraries called ncclient~\cite{ncclient} and paramiko~\cite{paramiko}. The callback subscriber and converters were implemented by utilizing Sysrepo-Python bindings~\cite{sysrepo-python} to subscribe to the datastore. Our translator supports OCSes from four different vendors; they could be controlled and monitored in a unified manner on the basis of our defined SBI.
	
	We deployed our implemented SDN controller on the general hardware platform of Intel(R) Xeon(R) Silver 4314 CPU @ 2.40GHz with 32 processors and 64GB of memory. The number of maximum worker threads for concurrent configurations of OCSes was set to 160 (= 32 * 5) in accordance with the default value explained in~\cite{concurrent}.

	\subsection{Evaluation on SBI}
	\label{subsec:exp-sbi}
	We first evaluated how our defined SBI helps operators reduce development cost. To this end, we evaluated the SLOC of the OCS driver, which was implemented inside our controller for configuring OCSes from four different vendors. As shown in Fig.~\ref{fig:driver-code-lines}, the SLOC can be significantly reduced with our SBI. Specifically, the SLOC was 153 lines with our SBI and 650 lines without it. Our SBI thus reduces SLOC by $76.4\%$. This result indicates the validity of defining the unified SBI for managing MV-OCSes. Note that the SLOC reduction could be significant as the number of supported MV-OCSes increases, which also indicates the validity of the unified SBI.
	
\begin{figure}[t]
		\centering
		\begin{tabular}{cc}
			{\begin{tabular}{@{}c@{}}
					\subfloat[Connection create operation]
					{\begin{tikzpicture}[gnuplot]
\tikzset{every node/.append style={font={\fontsize{7.0pt}{8.4pt}\selectfont}}}
\path (0.000,0.000) rectangle (4.572,3.810);
\gpcolor{color=gp lt color axes}
\gpsetlinetype{gp lt axes}
\gpsetdashtype{gp dt axes}
\gpsetlinewidth{0.50}
\draw[gp path] (1.054,0.691)--(4.184,0.691);
\gpcolor{color=gp lt color border}
\gpsetlinetype{gp lt border}
\gpsetdashtype{gp dt solid}
\gpsetlinewidth{1.00}
\draw[gp path] (1.054,0.691)--(1.234,0.691);
\draw[gp path] (4.184,0.691)--(4.004,0.691);
\node[gp node right] at (0.925,0.691) {$0$};
\gpcolor{color=gp lt color axes}
\gpsetlinetype{gp lt axes}
\gpsetdashtype{gp dt axes}
\gpsetlinewidth{0.50}
\draw[gp path] (1.054,1.054)--(4.184,1.054);
\gpcolor{color=gp lt color border}
\gpsetlinetype{gp lt border}
\gpsetdashtype{gp dt solid}
\gpsetlinewidth{1.00}
\draw[gp path] (1.054,1.054)--(1.234,1.054);
\draw[gp path] (4.184,1.054)--(4.004,1.054);
\node[gp node right] at (0.925,1.054) {$0.05$};
\gpcolor{color=gp lt color axes}
\gpsetlinetype{gp lt axes}
\gpsetdashtype{gp dt axes}
\gpsetlinewidth{0.50}
\draw[gp path] (1.054,1.417)--(4.184,1.417);
\gpcolor{color=gp lt color border}
\gpsetlinetype{gp lt border}
\gpsetdashtype{gp dt solid}
\gpsetlinewidth{1.00}
\draw[gp path] (1.054,1.417)--(1.234,1.417);
\draw[gp path] (4.184,1.417)--(4.004,1.417);
\node[gp node right] at (0.925,1.417) {$0.1$};
\gpcolor{color=gp lt color axes}
\gpsetlinetype{gp lt axes}
\gpsetdashtype{gp dt axes}
\gpsetlinewidth{0.50}
\draw[gp path] (1.054,1.779)--(4.184,1.779);
\gpcolor{color=gp lt color border}
\gpsetlinetype{gp lt border}
\gpsetdashtype{gp dt solid}
\gpsetlinewidth{1.00}
\draw[gp path] (1.054,1.779)--(1.234,1.779);
\draw[gp path] (4.184,1.779)--(4.004,1.779);
\node[gp node right] at (0.925,1.779) {$0.15$};
\gpcolor{color=gp lt color axes}
\gpsetlinetype{gp lt axes}
\gpsetdashtype{gp dt axes}
\gpsetlinewidth{0.50}
\draw[gp path] (1.054,2.142)--(4.184,2.142);
\gpcolor{color=gp lt color border}
\gpsetlinetype{gp lt border}
\gpsetdashtype{gp dt solid}
\gpsetlinewidth{1.00}
\draw[gp path] (1.054,2.142)--(1.234,2.142);
\draw[gp path] (4.184,2.142)--(4.004,2.142);
\node[gp node right] at (0.925,2.142) {$0.2$};
\gpcolor{color=gp lt color axes}
\gpsetlinetype{gp lt axes}
\gpsetdashtype{gp dt axes}
\gpsetlinewidth{0.50}
\draw[gp path] (1.054,2.505)--(4.184,2.505);
\gpcolor{color=gp lt color border}
\gpsetlinetype{gp lt border}
\gpsetdashtype{gp dt solid}
\gpsetlinewidth{1.00}
\draw[gp path] (1.054,2.505)--(1.234,2.505);
\draw[gp path] (4.184,2.505)--(4.004,2.505);
\node[gp node right] at (0.925,2.505) {$0.25$};
\gpcolor{color=gp lt color axes}
\gpsetlinetype{gp lt axes}
\gpsetdashtype{gp dt axes}
\gpsetlinewidth{0.50}
\draw[gp path] (1.054,2.868)--(2.069,2.868);
\draw[gp path] (4.055,2.868)--(4.184,2.868);
\gpcolor{color=gp lt color border}
\gpsetlinetype{gp lt border}
\gpsetdashtype{gp dt solid}
\gpsetlinewidth{1.00}
\draw[gp path] (1.054,2.868)--(1.234,2.868);
\draw[gp path] (4.184,2.868)--(4.004,2.868);
\node[gp node right] at (0.925,2.868) {$0.3$};
\gpcolor{color=gp lt color axes}
\gpsetlinetype{gp lt axes}
\gpsetdashtype{gp dt axes}
\gpsetlinewidth{0.50}
\draw[gp path] (1.054,3.230)--(2.069,3.230);
\draw[gp path] (4.055,3.230)--(4.184,3.230);
\gpcolor{color=gp lt color border}
\gpsetlinetype{gp lt border}
\gpsetdashtype{gp dt solid}
\gpsetlinewidth{1.00}
\draw[gp path] (1.054,3.230)--(1.234,3.230);
\draw[gp path] (4.184,3.230)--(4.004,3.230);
\node[gp node right] at (0.925,3.230) {$0.35$};
\gpcolor{color=gp lt color axes}
\gpsetlinetype{gp lt axes}
\gpsetdashtype{gp dt axes}
\gpsetlinewidth{0.50}
\draw[gp path] (1.054,3.593)--(4.184,3.593);
\gpcolor{color=gp lt color border}
\gpsetlinetype{gp lt border}
\gpsetdashtype{gp dt solid}
\gpsetlinewidth{1.00}
\draw[gp path] (1.054,3.593)--(1.234,3.593);
\draw[gp path] (4.184,3.593)--(4.004,3.593);
\node[gp node right] at (0.925,3.593) {$0.4$};
\gpcolor{color=gp lt color axes}
\gpsetlinetype{gp lt axes}
\gpsetdashtype{gp dt axes}
\gpsetlinewidth{0.50}
\draw[gp path] (1.198,0.691)--(1.198,3.593);
\gpcolor{color=gp lt color border}
\gpsetlinetype{gp lt border}
\gpsetdashtype{gp dt solid}
\gpsetlinewidth{1.00}
\draw[gp path] (1.198,0.691)--(1.198,0.871);
\draw[gp path] (1.198,3.593)--(1.198,3.413);
\node[gp node center] at (1.198,0.475) {1};
\gpcolor{color=gp lt color axes}
\gpsetlinetype{gp lt axes}
\gpsetdashtype{gp dt axes}
\gpsetlinewidth{0.50}
\draw[gp path] (2.145,0.691)--(2.145,2.738);
\draw[gp path] (2.145,3.413)--(2.145,3.593);
\gpcolor{color=gp lt color border}
\gpsetlinetype{gp lt border}
\gpsetdashtype{gp dt solid}
\gpsetlinewidth{1.00}
\draw[gp path] (2.145,0.691)--(2.145,0.871);
\draw[gp path] (2.145,3.593)--(2.145,3.413);
\node[gp node center] at (2.145,0.475) {2};
\gpcolor{color=gp lt color axes}
\gpsetlinetype{gp lt axes}
\gpsetdashtype{gp dt axes}
\gpsetlinewidth{0.50}
\draw[gp path] (3.092,0.691)--(3.092,2.738);
\draw[gp path] (3.092,3.413)--(3.092,3.593);
\gpcolor{color=gp lt color border}
\gpsetlinetype{gp lt border}
\gpsetdashtype{gp dt solid}
\gpsetlinewidth{1.00}
\draw[gp path] (3.092,0.691)--(3.092,0.871);
\draw[gp path] (3.092,3.593)--(3.092,3.413);
\node[gp node center] at (3.092,0.475) {4};
\gpcolor{color=gp lt color axes}
\gpsetlinetype{gp lt axes}
\gpsetdashtype{gp dt axes}
\gpsetlinewidth{0.50}
\draw[gp path] (4.038,0.691)--(4.038,2.738);
\draw[gp path] (4.038,3.413)--(4.038,3.593);
\gpcolor{color=gp lt color border}
\gpsetlinetype{gp lt border}
\gpsetdashtype{gp dt solid}
\gpsetlinewidth{1.00}
\draw[gp path] (4.038,0.691)--(4.038,0.871);
\draw[gp path] (4.038,3.593)--(4.038,3.413);
\node[gp node center] at (4.038,0.475) {8};
\draw[gp path] (1.054,3.593)--(1.054,0.691)--(4.184,0.691)--(4.184,3.593)--cycle;
\node[gp node center,rotate=-270] at (0.204,2.142) {Overhead [sec]};
\node[gp node center] at (2.619,0.151) {Num. of connection pairs};
\node[gp node right] at (3.101,3.300) {Vendor A};
\gpcolor{rgb color={0.580,0.000,0.827}}
\draw[gp path] (3.230,3.300)--(3.926,3.300);
\draw[gp path] (3.230,3.390)--(3.230,3.210);
\draw[gp path] (3.926,3.390)--(3.926,3.210);
\draw[gp path] (1.198,1.553)--(2.145,1.604)--(3.092,1.884)--(4.038,2.024);
\draw[gp path] (1.198,1.386)--(1.198,1.720);
\draw[gp path] (1.108,1.386)--(1.288,1.386);
\draw[gp path] (1.108,1.720)--(1.288,1.720);
\draw[gp path] (2.145,1.387)--(2.145,1.822);
\draw[gp path] (2.055,1.387)--(2.235,1.387);
\draw[gp path] (2.055,1.822)--(2.235,1.822);
\draw[gp path] (3.092,1.551)--(3.092,2.218);
\draw[gp path] (3.002,1.551)--(3.182,1.551);
\draw[gp path] (3.002,2.218)--(3.182,2.218);
\draw[gp path] (4.038,1.473)--(4.038,2.576);
\draw[gp path] (3.948,1.473)--(4.128,1.473);
\draw[gp path] (3.948,2.576)--(4.128,2.576);
\gpsetpointsize{4.00}
\gppoint{gp mark 6}{(1.198,1.553)}
\gppoint{gp mark 6}{(2.145,1.604)}
\gppoint{gp mark 6}{(3.092,1.884)}
\gppoint{gp mark 6}{(4.038,2.024)}
\gppoint{gp mark 6}{(3.578,3.300)}
\gpcolor{color=gp lt color border}
\node[gp node right] at (3.101,3.075) {Vendor B};
\gpcolor{rgb color={0.000,0.620,0.451}}
\draw[gp path] (3.230,3.075)--(3.926,3.075);
\draw[gp path] (3.230,3.165)--(3.230,2.985);
\draw[gp path] (3.926,3.165)--(3.926,2.985);
\draw[gp path] (1.198,1.593)--(2.145,1.884)--(3.092,2.035)--(4.038,2.190);
\draw[gp path] (1.198,1.295)--(1.198,1.890);
\draw[gp path] (1.108,1.295)--(1.288,1.295);
\draw[gp path] (1.108,1.890)--(1.288,1.890);
\draw[gp path] (2.145,1.442)--(2.145,2.327);
\draw[gp path] (2.055,1.442)--(2.235,1.442);
\draw[gp path] (2.055,2.327)--(2.235,2.327);
\draw[gp path] (3.092,1.621)--(3.092,2.448);
\draw[gp path] (3.002,1.621)--(3.182,1.621);
\draw[gp path] (3.002,2.448)--(3.182,2.448);
\draw[gp path] (4.038,1.755)--(4.038,2.625);
\draw[gp path] (3.948,1.755)--(4.128,1.755);
\draw[gp path] (3.948,2.625)--(4.128,2.625);
\gppoint{gp mark 7}{(1.198,1.593)}
\gppoint{gp mark 7}{(2.145,1.884)}
\gppoint{gp mark 7}{(3.092,2.035)}
\gppoint{gp mark 7}{(4.038,2.190)}
\gppoint{gp mark 7}{(3.578,3.075)}
\gpcolor{color=gp lt color border}
\node[gp node right] at (3.101,2.850) {Vendor C};
\gpcolor{rgb color={0.337,0.706,0.914}}
\draw[gp path] (3.230,2.850)--(3.926,2.850);
\draw[gp path] (3.230,2.940)--(3.230,2.760);
\draw[gp path] (3.926,2.940)--(3.926,2.760);
\draw[gp path] (1.198,1.520)--(2.145,1.728)--(3.092,1.808)--(4.038,2.005);
\draw[gp path] (1.198,1.200)--(1.198,1.839);
\draw[gp path] (1.108,1.200)--(1.288,1.200);
\draw[gp path] (1.108,1.839)--(1.288,1.839);
\draw[gp path] (2.145,1.518)--(2.145,1.939);
\draw[gp path] (2.055,1.518)--(2.235,1.518);
\draw[gp path] (2.055,1.939)--(2.235,1.939);
\draw[gp path] (3.092,1.467)--(3.092,2.149);
\draw[gp path] (3.002,1.467)--(3.182,1.467);
\draw[gp path] (3.002,2.149)--(3.182,2.149);
\draw[gp path] (4.038,1.613)--(4.038,2.397);
\draw[gp path] (3.948,1.613)--(4.128,1.613);
\draw[gp path] (3.948,2.397)--(4.128,2.397);
\gppoint{gp mark 8}{(1.198,1.520)}
\gppoint{gp mark 8}{(2.145,1.728)}
\gppoint{gp mark 8}{(3.092,1.808)}
\gppoint{gp mark 8}{(4.038,2.005)}
\gppoint{gp mark 8}{(3.578,2.850)}
\gpcolor{color=gp lt color border}
\draw[gp path] (1.054,3.593)--(1.054,0.691)--(4.184,0.691)--(4.184,3.593)--cycle;
\gpdefrectangularnode{gp plot 1}{\pgfpoint{1.054cm}{0.691cm}}{\pgfpoint{4.184cm}{3.593cm}}
\end{tikzpicture}
}
					\subfloat[Connection delete operation]
					{\begin{tikzpicture}[gnuplot]
\tikzset{every node/.append style={font={\fontsize{7.0pt}{8.4pt}\selectfont}}}
\path (0.000,0.000) rectangle (4.318,3.810);
\gpcolor{color=gp lt color axes}
\gpsetlinetype{gp lt axes}
\gpsetdashtype{gp dt axes}
\gpsetlinewidth{0.50}
\draw[gp path] (0.838,0.691)--(3.930,0.691);
\gpcolor{color=gp lt color border}
\gpsetlinetype{gp lt border}
\gpsetdashtype{gp dt solid}
\gpsetlinewidth{1.00}
\draw[gp path] (0.838,0.691)--(1.018,0.691);
\draw[gp path] (3.930,0.691)--(3.750,0.691);
\node[gp node right] at (0.709,0.691) {$0$};
\gpcolor{color=gp lt color axes}
\gpsetlinetype{gp lt axes}
\gpsetdashtype{gp dt axes}
\gpsetlinewidth{0.50}
\draw[gp path] (0.838,1.054)--(3.930,1.054);
\gpcolor{color=gp lt color border}
\gpsetlinetype{gp lt border}
\gpsetdashtype{gp dt solid}
\gpsetlinewidth{1.00}
\draw[gp path] (0.838,1.054)--(1.018,1.054);
\draw[gp path] (3.930,1.054)--(3.750,1.054);
\node[gp node right] at (0.709,1.054) {$0.05$};
\gpcolor{color=gp lt color axes}
\gpsetlinetype{gp lt axes}
\gpsetdashtype{gp dt axes}
\gpsetlinewidth{0.50}
\draw[gp path] (0.838,1.417)--(3.930,1.417);
\gpcolor{color=gp lt color border}
\gpsetlinetype{gp lt border}
\gpsetdashtype{gp dt solid}
\gpsetlinewidth{1.00}
\draw[gp path] (0.838,1.417)--(1.018,1.417);
\draw[gp path] (3.930,1.417)--(3.750,1.417);
\node[gp node right] at (0.709,1.417) {$0.1$};
\gpcolor{color=gp lt color axes}
\gpsetlinetype{gp lt axes}
\gpsetdashtype{gp dt axes}
\gpsetlinewidth{0.50}
\draw[gp path] (0.838,1.779)--(3.930,1.779);
\gpcolor{color=gp lt color border}
\gpsetlinetype{gp lt border}
\gpsetdashtype{gp dt solid}
\gpsetlinewidth{1.00}
\draw[gp path] (0.838,1.779)--(1.018,1.779);
\draw[gp path] (3.930,1.779)--(3.750,1.779);
\node[gp node right] at (0.709,1.779) {$0.15$};
\gpcolor{color=gp lt color axes}
\gpsetlinetype{gp lt axes}
\gpsetdashtype{gp dt axes}
\gpsetlinewidth{0.50}
\draw[gp path] (0.838,2.142)--(3.930,2.142);
\gpcolor{color=gp lt color border}
\gpsetlinetype{gp lt border}
\gpsetdashtype{gp dt solid}
\gpsetlinewidth{1.00}
\draw[gp path] (0.838,2.142)--(1.018,2.142);
\draw[gp path] (3.930,2.142)--(3.750,2.142);
\node[gp node right] at (0.709,2.142) {$0.2$};
\gpcolor{color=gp lt color axes}
\gpsetlinetype{gp lt axes}
\gpsetdashtype{gp dt axes}
\gpsetlinewidth{0.50}
\draw[gp path] (0.838,2.505)--(3.930,2.505);
\gpcolor{color=gp lt color border}
\gpsetlinetype{gp lt border}
\gpsetdashtype{gp dt solid}
\gpsetlinewidth{1.00}
\draw[gp path] (0.838,2.505)--(1.018,2.505);
\draw[gp path] (3.930,2.505)--(3.750,2.505);
\node[gp node right] at (0.709,2.505) {$0.25$};
\gpcolor{color=gp lt color axes}
\gpsetlinetype{gp lt axes}
\gpsetdashtype{gp dt axes}
\gpsetlinewidth{0.50}
\draw[gp path] (0.838,2.868)--(1.815,2.868);
\draw[gp path] (3.801,2.868)--(3.930,2.868);
\gpcolor{color=gp lt color border}
\gpsetlinetype{gp lt border}
\gpsetdashtype{gp dt solid}
\gpsetlinewidth{1.00}
\draw[gp path] (0.838,2.868)--(1.018,2.868);
\draw[gp path] (3.930,2.868)--(3.750,2.868);
\node[gp node right] at (0.709,2.868) {$0.3$};
\gpcolor{color=gp lt color axes}
\gpsetlinetype{gp lt axes}
\gpsetdashtype{gp dt axes}
\gpsetlinewidth{0.50}
\draw[gp path] (0.838,3.230)--(1.815,3.230);
\draw[gp path] (3.801,3.230)--(3.930,3.230);
\gpcolor{color=gp lt color border}
\gpsetlinetype{gp lt border}
\gpsetdashtype{gp dt solid}
\gpsetlinewidth{1.00}
\draw[gp path] (0.838,3.230)--(1.018,3.230);
\draw[gp path] (3.930,3.230)--(3.750,3.230);
\node[gp node right] at (0.709,3.230) {$0.35$};
\gpcolor{color=gp lt color axes}
\gpsetlinetype{gp lt axes}
\gpsetdashtype{gp dt axes}
\gpsetlinewidth{0.50}
\draw[gp path] (0.838,3.593)--(3.930,3.593);
\gpcolor{color=gp lt color border}
\gpsetlinetype{gp lt border}
\gpsetdashtype{gp dt solid}
\gpsetlinewidth{1.00}
\draw[gp path] (0.838,3.593)--(1.018,3.593);
\draw[gp path] (3.930,3.593)--(3.750,3.593);
\node[gp node right] at (0.709,3.593) {$0.4$};
\gpcolor{color=gp lt color axes}
\gpsetlinetype{gp lt axes}
\gpsetdashtype{gp dt axes}
\gpsetlinewidth{0.50}
\draw[gp path] (0.980,0.691)--(0.980,3.593);
\gpcolor{color=gp lt color border}
\gpsetlinetype{gp lt border}
\gpsetdashtype{gp dt solid}
\gpsetlinewidth{1.00}
\draw[gp path] (0.980,0.691)--(0.980,0.871);
\draw[gp path] (0.980,3.593)--(0.980,3.413);
\node[gp node center] at (0.980,0.475) {1};
\gpcolor{color=gp lt color axes}
\gpsetlinetype{gp lt axes}
\gpsetdashtype{gp dt axes}
\gpsetlinewidth{0.50}
\draw[gp path] (1.915,0.691)--(1.915,2.738);
\draw[gp path] (1.915,3.413)--(1.915,3.593);
\gpcolor{color=gp lt color border}
\gpsetlinetype{gp lt border}
\gpsetdashtype{gp dt solid}
\gpsetlinewidth{1.00}
\draw[gp path] (1.915,0.691)--(1.915,0.871);
\draw[gp path] (1.915,3.593)--(1.915,3.413);
\node[gp node center] at (1.915,0.475) {2};
\gpcolor{color=gp lt color axes}
\gpsetlinetype{gp lt axes}
\gpsetdashtype{gp dt axes}
\gpsetlinewidth{0.50}
\draw[gp path] (2.851,0.691)--(2.851,2.738);
\draw[gp path] (2.851,3.413)--(2.851,3.593);
\gpcolor{color=gp lt color border}
\gpsetlinetype{gp lt border}
\gpsetdashtype{gp dt solid}
\gpsetlinewidth{1.00}
\draw[gp path] (2.851,0.691)--(2.851,0.871);
\draw[gp path] (2.851,3.593)--(2.851,3.413);
\node[gp node center] at (2.851,0.475) {4};
\gpcolor{color=gp lt color axes}
\gpsetlinetype{gp lt axes}
\gpsetdashtype{gp dt axes}
\gpsetlinewidth{0.50}
\draw[gp path] (3.786,0.691)--(3.786,2.738);
\draw[gp path] (3.786,3.413)--(3.786,3.593);
\gpcolor{color=gp lt color border}
\gpsetlinetype{gp lt border}
\gpsetdashtype{gp dt solid}
\gpsetlinewidth{1.00}
\draw[gp path] (3.786,0.691)--(3.786,0.871);
\draw[gp path] (3.786,3.593)--(3.786,3.413);
\node[gp node center] at (3.786,0.475) {8};
\draw[gp path] (0.838,3.593)--(0.838,0.691)--(3.930,0.691)--(3.930,3.593)--cycle;
\node[gp node center] at (2.384,0.151) {Num. of connection pairs};
\node[gp node right] at (2.847,3.300) {Vendor A};
\gpcolor{rgb color={0.580,0.000,0.827}}
\draw[gp path] (2.976,3.300)--(3.672,3.300);
\draw[gp path] (2.976,3.390)--(2.976,3.210);
\draw[gp path] (3.672,3.390)--(3.672,3.210);
\draw[gp path] (0.980,1.821)--(1.915,1.898)--(2.851,1.987)--(3.786,2.075);
\draw[gp path] (0.980,1.560)--(0.980,2.083);
\draw[gp path] (0.890,1.560)--(1.070,1.560);
\draw[gp path] (0.890,2.083)--(1.070,2.083);
\draw[gp path] (1.915,1.681)--(1.915,2.116);
\draw[gp path] (1.825,1.681)--(2.005,1.681);
\draw[gp path] (1.825,2.116)--(2.005,2.116);
\draw[gp path] (2.851,1.835)--(2.851,2.140);
\draw[gp path] (2.761,1.835)--(2.941,1.835);
\draw[gp path] (2.761,2.140)--(2.941,2.140);
\draw[gp path] (3.786,1.792)--(3.786,2.357);
\draw[gp path] (3.696,1.792)--(3.876,1.792);
\draw[gp path] (3.696,2.357)--(3.876,2.357);
\gpsetpointsize{4.00}
\gppoint{gp mark 6}{(0.980,1.821)}
\gppoint{gp mark 6}{(1.915,1.898)}
\gppoint{gp mark 6}{(2.851,1.987)}
\gppoint{gp mark 6}{(3.786,2.075)}
\gppoint{gp mark 6}{(3.324,3.300)}
\gpcolor{color=gp lt color border}
\node[gp node right] at (2.847,3.075) {Vendor B};
\gpcolor{rgb color={0.000,0.620,0.451}}
\draw[gp path] (2.976,3.075)--(3.672,3.075);
\draw[gp path] (2.976,3.165)--(2.976,2.985);
\draw[gp path] (3.672,3.165)--(3.672,2.985);
\draw[gp path] (0.980,1.431)--(1.915,1.562)--(2.851,1.767)--(3.786,1.879);
\draw[gp path] (0.980,1.141)--(0.980,1.721);
\draw[gp path] (0.890,1.141)--(1.070,1.141);
\draw[gp path] (0.890,1.721)--(1.070,1.721);
\draw[gp path] (1.915,1.395)--(1.915,1.729);
\draw[gp path] (1.825,1.395)--(2.005,1.395);
\draw[gp path] (1.825,1.729)--(2.005,1.729);
\draw[gp path] (2.851,1.477)--(2.851,2.057);
\draw[gp path] (2.761,1.477)--(2.941,1.477);
\draw[gp path] (2.761,2.057)--(2.941,2.057);
\draw[gp path] (3.786,1.625)--(3.786,2.133);
\draw[gp path] (3.696,1.625)--(3.876,1.625);
\draw[gp path] (3.696,2.133)--(3.876,2.133);
\gppoint{gp mark 7}{(0.980,1.431)}
\gppoint{gp mark 7}{(1.915,1.562)}
\gppoint{gp mark 7}{(2.851,1.767)}
\gppoint{gp mark 7}{(3.786,1.879)}
\gppoint{gp mark 7}{(3.324,3.075)}
\gpcolor{color=gp lt color border}
\node[gp node right] at (2.847,2.850) {Vendor C};
\gpcolor{rgb color={0.337,0.706,0.914}}
\draw[gp path] (2.976,2.850)--(3.672,2.850);
\draw[gp path] (2.976,2.940)--(2.976,2.760);
\draw[gp path] (3.672,2.940)--(3.672,2.760);
\draw[gp path] (0.980,1.461)--(1.915,1.539)--(2.851,1.915)--(3.786,1.990);
\draw[gp path] (0.980,1.208)--(0.980,1.715);
\draw[gp path] (0.890,1.208)--(1.070,1.208);
\draw[gp path] (0.890,1.715)--(1.070,1.715);
\draw[gp path] (1.915,1.213)--(1.915,1.866);
\draw[gp path] (1.825,1.213)--(2.005,1.213);
\draw[gp path] (1.825,1.866)--(2.005,1.866);
\draw[gp path] (2.851,1.487)--(2.851,2.343);
\draw[gp path] (2.761,1.487)--(2.941,1.487);
\draw[gp path] (2.761,2.343)--(2.941,2.343);
\draw[gp path] (3.786,1.540)--(3.786,2.439);
\draw[gp path] (3.696,1.540)--(3.876,1.540);
\draw[gp path] (3.696,2.439)--(3.876,2.439);
\gppoint{gp mark 8}{(0.980,1.461)}
\gppoint{gp mark 8}{(1.915,1.539)}
\gppoint{gp mark 8}{(2.851,1.915)}
\gppoint{gp mark 8}{(3.786,1.990)}
\gppoint{gp mark 8}{(3.324,2.850)}
\gpcolor{color=gp lt color border}
\draw[gp path] (0.838,3.593)--(0.838,0.691)--(3.930,0.691)--(3.930,3.593)--cycle;
\gpdefrectangularnode{gp plot 1}{\pgfpoint{0.838cm}{0.691cm}}{\pgfpoint{3.930cm}{3.593cm}}
\end{tikzpicture}
}
			\end{tabular}}
		\end{tabular}
		\caption{Overhead incurred by our SBI vs. number of configured internal connections.}
		\label{fig:results-overhead}
	\end{figure}
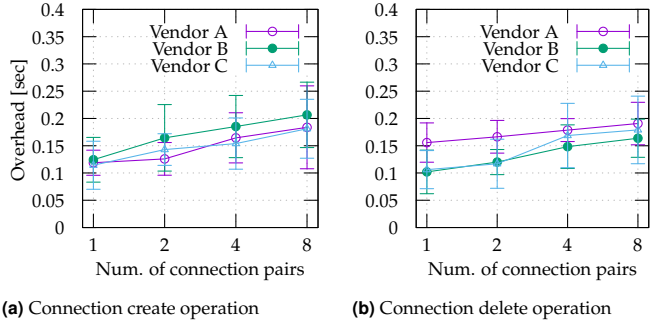

	Second, we evaluated the overhead incurred by our SBI for creating and deleting different numbers of internal connections on OCSes. Operating MV-OCSes with our SBI incurs additional configuration time compared with using vendor-proprietary APIs due to translation between NETCONF operations on OCS.yang and vendor-proprietary APIs. Thus, we experimentally studied how large the overhead was. We evaluated it by executing internal connection create and delete operations with our SBI $10$ times and show the averaged values with standard deviations. As shown in Fig.~\ref{fig:results-overhead}, the overhead tended to very slightly increase as the number of internal connections increased for all vendors. Since Extensible Markup Language (XML) payload becomes large as the number of internal connection configurations increases, the time to parse and convert it to the vendor-proprietary commands also increases. However, we found that the overhead did not exceed $0.30$ seconds under any settings. The overhead was less than the configuration time of vendor-proprietary APIs and accounted for small portions of overall configuration time of OCSes; it was at most $36.8\%$ on all OCSes. From the results, we found that requirement~\ref{sec:req}.\ref{subsec:req-b} was satisfied. Here, the raw data of configuration time of each vendor's OCS is not provided because it is confidential information.
	
	\begin{figure}[t]
		\centering
		\begin{tikzpicture}[gnuplot]
\tikzset{every node/.append style={font={\fontsize{7.0pt}{8.4pt}\selectfont}}}
\path (0.000,0.000) rectangle (6.350,3.810);
\gpcolor{color=gp lt color axes}
\gpsetlinetype{gp lt axes}
\gpsetdashtype{gp dt axes}
\gpsetlinewidth{0.50}
\draw[gp path] (0.925,0.432)--(5.962,0.432);
\gpcolor{color=gp lt color border}
\gpsetlinetype{gp lt border}
\gpsetdashtype{gp dt solid}
\gpsetlinewidth{1.00}
\draw[gp path] (0.925,0.432)--(1.105,0.432);
\draw[gp path] (5.962,0.432)--(5.782,0.432);
\node[gp node right] at (0.796,0.432) {$0$};
\gpcolor{color=gp lt color axes}
\gpsetlinetype{gp lt axes}
\gpsetdashtype{gp dt axes}
\gpsetlinewidth{0.50}
\draw[gp path] (0.925,0.827)--(5.962,0.827);
\gpcolor{color=gp lt color border}
\gpsetlinetype{gp lt border}
\gpsetdashtype{gp dt solid}
\gpsetlinewidth{1.00}
\draw[gp path] (0.925,0.827)--(1.105,0.827);
\draw[gp path] (5.962,0.827)--(5.782,0.827);
\node[gp node right] at (0.796,0.827) {$100$};
\gpcolor{color=gp lt color axes}
\gpsetlinetype{gp lt axes}
\gpsetdashtype{gp dt axes}
\gpsetlinewidth{0.50}
\draw[gp path] (0.925,1.222)--(5.962,1.222);
\gpcolor{color=gp lt color border}
\gpsetlinetype{gp lt border}
\gpsetdashtype{gp dt solid}
\gpsetlinewidth{1.00}
\draw[gp path] (0.925,1.222)--(1.105,1.222);
\draw[gp path] (5.962,1.222)--(5.782,1.222);
\node[gp node right] at (0.796,1.222) {$200$};
\gpcolor{color=gp lt color axes}
\gpsetlinetype{gp lt axes}
\gpsetdashtype{gp dt axes}
\gpsetlinewidth{0.50}
\draw[gp path] (0.925,1.617)--(5.962,1.617);
\gpcolor{color=gp lt color border}
\gpsetlinetype{gp lt border}
\gpsetdashtype{gp dt solid}
\gpsetlinewidth{1.00}
\draw[gp path] (0.925,1.617)--(1.105,1.617);
\draw[gp path] (5.962,1.617)--(5.782,1.617);
\node[gp node right] at (0.796,1.617) {$300$};
\gpcolor{color=gp lt color axes}
\gpsetlinetype{gp lt axes}
\gpsetdashtype{gp dt axes}
\gpsetlinewidth{0.50}
\draw[gp path] (0.925,2.013)--(5.962,2.013);
\gpcolor{color=gp lt color border}
\gpsetlinetype{gp lt border}
\gpsetdashtype{gp dt solid}
\gpsetlinewidth{1.00}
\draw[gp path] (0.925,2.013)--(1.105,2.013);
\draw[gp path] (5.962,2.013)--(5.782,2.013);
\node[gp node right] at (0.796,2.013) {$400$};
\gpcolor{color=gp lt color axes}
\gpsetlinetype{gp lt axes}
\gpsetdashtype{gp dt axes}
\gpsetlinewidth{0.50}
\draw[gp path] (0.925,2.408)--(5.962,2.408);
\gpcolor{color=gp lt color border}
\gpsetlinetype{gp lt border}
\gpsetdashtype{gp dt solid}
\gpsetlinewidth{1.00}
\draw[gp path] (0.925,2.408)--(1.105,2.408);
\draw[gp path] (5.962,2.408)--(5.782,2.408);
\node[gp node right] at (0.796,2.408) {$500$};
\gpcolor{color=gp lt color axes}
\gpsetlinetype{gp lt axes}
\gpsetdashtype{gp dt axes}
\gpsetlinewidth{0.50}
\draw[gp path] (0.925,2.803)--(5.962,2.803);
\gpcolor{color=gp lt color border}
\gpsetlinetype{gp lt border}
\gpsetdashtype{gp dt solid}
\gpsetlinewidth{1.00}
\draw[gp path] (0.925,2.803)--(1.105,2.803);
\draw[gp path] (5.962,2.803)--(5.782,2.803);
\node[gp node right] at (0.796,2.803) {$600$};
\gpcolor{color=gp lt color axes}
\gpsetlinetype{gp lt axes}
\gpsetdashtype{gp dt axes}
\gpsetlinewidth{0.50}
\draw[gp path] (0.925,3.198)--(5.962,3.198);
\gpcolor{color=gp lt color border}
\gpsetlinetype{gp lt border}
\gpsetdashtype{gp dt solid}
\gpsetlinewidth{1.00}
\draw[gp path] (0.925,3.198)--(1.105,3.198);
\draw[gp path] (5.962,3.198)--(5.782,3.198);
\node[gp node right] at (0.796,3.198) {$700$};
\gpcolor{color=gp lt color axes}
\gpsetlinetype{gp lt axes}
\gpsetdashtype{gp dt axes}
\gpsetlinewidth{0.50}
\draw[gp path] (0.925,3.593)--(5.962,3.593);
\gpcolor{color=gp lt color border}
\gpsetlinetype{gp lt border}
\gpsetdashtype{gp dt solid}
\gpsetlinewidth{1.00}
\draw[gp path] (0.925,3.593)--(1.105,3.593);
\draw[gp path] (5.962,3.593)--(5.782,3.593);
\node[gp node right] at (0.796,3.593) {$800$};
\gpcolor{color=gp lt color axes}
\gpsetlinetype{gp lt axes}
\gpsetdashtype{gp dt axes}
\gpsetlinewidth{0.50}
\draw[gp path] (2.184,0.432)--(2.184,3.593);
\gpcolor{color=gp lt color border}
\gpsetlinetype{gp lt border}
\gpsetdashtype{gp dt solid}
\gpsetlinewidth{1.00}
\draw[gp path] (2.184,0.432)--(2.184,0.612);
\draw[gp path] (2.184,3.593)--(2.184,3.413);
\node[gp node center] at (2.184,0.216) {w/-SBI};
\gpcolor{color=gp lt color axes}
\gpsetlinetype{gp lt axes}
\gpsetdashtype{gp dt axes}
\gpsetlinewidth{0.50}
\draw[gp path] (4.703,0.432)--(4.703,3.593);
\gpcolor{color=gp lt color border}
\gpsetlinetype{gp lt border}
\gpsetdashtype{gp dt solid}
\gpsetlinewidth{1.00}
\draw[gp path] (4.703,0.432)--(4.703,0.612);
\draw[gp path] (4.703,3.593)--(4.703,3.413);
\node[gp node center] at (4.703,0.216) {w/o-SBI};
\draw[gp path] (0.925,3.593)--(0.925,0.432)--(5.962,0.432)--(5.962,3.593)--cycle;
\node[gp node center,rotate=-270] at (0.204,2.012) {Lines};
\def\gpfillpath{(1.932,0.432)--(2.437,0.432)--(2.437,1.038)--(1.932,1.038)--cycle}
\gpfill{color=gpbgfillcolor} \gpfillpath;
\gpfill{rgb color={1.000,0.647,0.000},gp pattern 2,pattern color=.} \gpfillpath;
\gpcolor{rgb color={1.000,0.647,0.000}}
\draw[gp path] (1.932,0.432)--(1.932,1.037)--(2.436,1.037)--(2.436,0.432)--cycle;
\def\gpfillpath{(4.451,0.432)--(4.956,0.432)--(4.956,3.001)--(4.451,3.001)--cycle}
\gpfill{color=gpbgfillcolor} \gpfillpath;
\gpfill{rgb color={1.000,0.647,0.000},gp pattern 2,pattern color=.} \gpfillpath;
\draw[gp path] (4.451,0.432)--(4.451,3.000)--(4.955,3.000)--(4.955,0.432)--cycle;
\gpcolor{color=gp lt color border}
\draw[gp path] (0.925,3.593)--(0.925,0.432)--(5.962,0.432)--(5.962,3.593)--cycle;
\gpdefrectangularnode{gp plot 1}{\pgfpoint{0.925cm}{0.432cm}}{\pgfpoint{5.962cm}{3.593cm}}
\end{tikzpicture}
		\caption{The SLOC of OCS driver.}
		\label{fig:driver-code-lines}
	\end{figure}
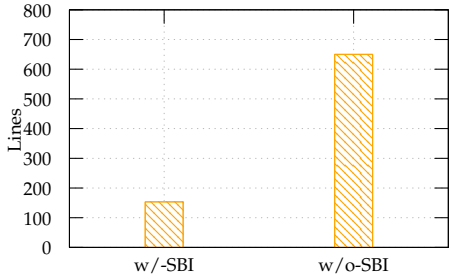
	
	\subsection{Evaluation on NBI}
	\label{subsec:exp-nbi}

\begin{figure}[t]
		\centering
		\includegraphics[width=.48\textwidth]{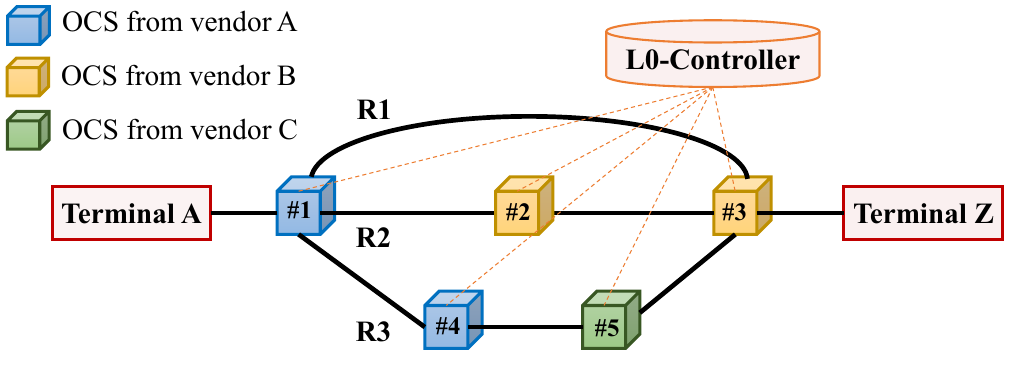}
		\caption{Experimental testbed with real MV-OCSes.}
		\label{fig:exp-env}
	\end{figure}

	We then evaluated the feasibility of our NBI as well as controller performance. We set up the experimental testbed using real MV-OCSes in our lab; it consists of five OCSes (OCSes \#1 to \# 5) from three different vendors and two terminals (terminals A and Z) as shown in Fig.~\ref{fig:exp-env}. There were three routes between terminals A and Z: R1, R2, and R3. We evaluated the completion time to establish or release the fiber paths on each route and present their averaged values with standard deviations. Here, the completion time is defined as the time required to complete the configurations for all OCSes to establish or release fiber paths by the controller. This could be checked by receiving the ACK from all of them. In addition, we confirmed the actual path setup and termination by checking the operational state of terminals (i.e., the state of link up or down). Though this evaluation focuses on the testbed configuration shown in Fig.~\ref{fig:exp-env}, we also previously found that our MV-OCS controller successfully operated the real OCS-based intra-DCNs in~\cite{anazawa2024first}.
	
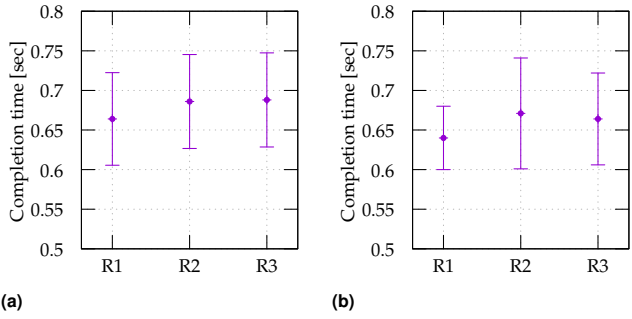
\begin{figure}[t]
		\centering
		\begin{tabular}{cc}
			{\begin{tabular}{@{}c@{}}
					\subfloat[]{\begin{tikzpicture}[gnuplot]
\tikzset{every node/.append style={font={\fontsize{7.0pt}{8.4pt}\selectfont}}}
\path (0.000,0.000) rectangle (4.318,3.810);
\gpcolor{color=gp lt color axes}
\gpsetlinetype{gp lt axes}
\gpsetdashtype{gp dt axes}
\gpsetlinewidth{0.50}
\draw[gp path] (1.054,0.432)--(3.930,0.432);
\gpcolor{color=gp lt color border}
\gpsetlinetype{gp lt border}
\gpsetdashtype{gp dt solid}
\gpsetlinewidth{1.00}
\draw[gp path] (1.054,0.432)--(1.234,0.432);
\draw[gp path] (3.930,0.432)--(3.750,0.432);
\node[gp node right] at (0.925,0.432) {$0.5$};
\gpcolor{color=gp lt color axes}
\gpsetlinetype{gp lt axes}
\gpsetdashtype{gp dt axes}
\gpsetlinewidth{0.50}
\draw[gp path] (1.054,0.959)--(3.930,0.959);
\gpcolor{color=gp lt color border}
\gpsetlinetype{gp lt border}
\gpsetdashtype{gp dt solid}
\gpsetlinewidth{1.00}
\draw[gp path] (1.054,0.959)--(1.234,0.959);
\draw[gp path] (3.930,0.959)--(3.750,0.959);
\node[gp node right] at (0.925,0.959) {$0.55$};
\gpcolor{color=gp lt color axes}
\gpsetlinetype{gp lt axes}
\gpsetdashtype{gp dt axes}
\gpsetlinewidth{0.50}
\draw[gp path] (1.054,1.486)--(3.930,1.486);
\gpcolor{color=gp lt color border}
\gpsetlinetype{gp lt border}
\gpsetdashtype{gp dt solid}
\gpsetlinewidth{1.00}
\draw[gp path] (1.054,1.486)--(1.234,1.486);
\draw[gp path] (3.930,1.486)--(3.750,1.486);
\node[gp node right] at (0.925,1.486) {$0.6$};
\gpcolor{color=gp lt color axes}
\gpsetlinetype{gp lt axes}
\gpsetdashtype{gp dt axes}
\gpsetlinewidth{0.50}
\draw[gp path] (1.054,2.013)--(3.930,2.013);
\gpcolor{color=gp lt color border}
\gpsetlinetype{gp lt border}
\gpsetdashtype{gp dt solid}
\gpsetlinewidth{1.00}
\draw[gp path] (1.054,2.013)--(1.234,2.013);
\draw[gp path] (3.930,2.013)--(3.750,2.013);
\node[gp node right] at (0.925,2.013) {$0.65$};
\gpcolor{color=gp lt color axes}
\gpsetlinetype{gp lt axes}
\gpsetdashtype{gp dt axes}
\gpsetlinewidth{0.50}
\draw[gp path] (1.054,2.539)--(3.930,2.539);
\gpcolor{color=gp lt color border}
\gpsetlinetype{gp lt border}
\gpsetdashtype{gp dt solid}
\gpsetlinewidth{1.00}
\draw[gp path] (1.054,2.539)--(1.234,2.539);
\draw[gp path] (3.930,2.539)--(3.750,2.539);
\node[gp node right] at (0.925,2.539) {$0.7$};
\gpcolor{color=gp lt color axes}
\gpsetlinetype{gp lt axes}
\gpsetdashtype{gp dt axes}
\gpsetlinewidth{0.50}
\draw[gp path] (1.054,3.066)--(3.930,3.066);
\gpcolor{color=gp lt color border}
\gpsetlinetype{gp lt border}
\gpsetdashtype{gp dt solid}
\gpsetlinewidth{1.00}
\draw[gp path] (1.054,3.066)--(1.234,3.066);
\draw[gp path] (3.930,3.066)--(3.750,3.066);
\node[gp node right] at (0.925,3.066) {$0.75$};
\gpcolor{color=gp lt color axes}
\gpsetlinetype{gp lt axes}
\gpsetdashtype{gp dt axes}
\gpsetlinewidth{0.50}
\draw[gp path] (1.054,3.593)--(3.930,3.593);
\gpcolor{color=gp lt color border}
\gpsetlinetype{gp lt border}
\gpsetdashtype{gp dt solid}
\gpsetlinewidth{1.00}
\draw[gp path] (1.054,3.593)--(1.234,3.593);
\draw[gp path] (3.930,3.593)--(3.750,3.593);
\node[gp node right] at (0.925,3.593) {$0.8$};
\gpcolor{color=gp lt color axes}
\gpsetlinetype{gp lt axes}
\gpsetdashtype{gp dt axes}
\gpsetlinewidth{0.50}
\draw[gp path] (1.465,0.432)--(1.465,3.593);
\gpcolor{color=gp lt color border}
\gpsetlinetype{gp lt border}
\gpsetdashtype{gp dt solid}
\gpsetlinewidth{1.00}
\draw[gp path] (1.465,0.432)--(1.465,0.612);
\draw[gp path] (1.465,3.593)--(1.465,3.413);
\node[gp node center] at (1.465,0.216) {R1};
\gpcolor{color=gp lt color axes}
\gpsetlinetype{gp lt axes}
\gpsetdashtype{gp dt axes}
\gpsetlinewidth{0.50}
\draw[gp path] (2.492,0.432)--(2.492,3.593);
\gpcolor{color=gp lt color border}
\gpsetlinetype{gp lt border}
\gpsetdashtype{gp dt solid}
\gpsetlinewidth{1.00}
\draw[gp path] (2.492,0.432)--(2.492,0.612);
\draw[gp path] (2.492,3.593)--(2.492,3.413);
\node[gp node center] at (2.492,0.216) {R2};
\gpcolor{color=gp lt color axes}
\gpsetlinetype{gp lt axes}
\gpsetdashtype{gp dt axes}
\gpsetlinewidth{0.50}
\draw[gp path] (3.519,0.432)--(3.519,3.413)--(3.519,3.593);
\gpcolor{color=gp lt color border}
\gpsetlinetype{gp lt border}
\gpsetdashtype{gp dt solid}
\gpsetlinewidth{1.00}
\draw[gp path] (3.519,0.432)--(3.519,0.612);
\draw[gp path] (3.519,3.593)--(3.519,3.413);
\node[gp node center] at (3.519,0.216) {R3};
\draw[gp path] (1.054,3.593)--(1.054,0.432)--(3.930,0.432)--(3.930,3.593)--cycle;
\node[gp node center,rotate=-270] at (0.204,2.012) {Completion time [sec]};
\gpcolor{rgb color={0.580,0.000,0.827}}
\draw[gp path] (1.465,1.544)--(1.465,2.776);
\draw[gp path] (1.375,1.544)--(1.555,1.544);
\draw[gp path] (1.375,2.776)--(1.555,2.776);
\draw[gp path] (2.492,1.767)--(2.492,3.017);
\draw[gp path] (2.402,1.767)--(2.582,1.767);
\draw[gp path] (2.402,3.017)--(2.582,3.017);
\draw[gp path] (3.519,1.787)--(3.519,3.039);
\draw[gp path] (3.429,1.787)--(3.609,1.787);
\draw[gp path] (3.429,3.039)--(3.609,3.039);
\gpsetpointsize{4.00}
\gppoint{gp mark 1}{(1.465,2.160)}
\gppoint{gp mark 1}{(2.492,2.392)}
\gppoint{gp mark 1}{(3.519,2.413)}
\gpsetpointsize{2.40}
\gppoint{gp mark 7}{(1.465,2.160)}
\gppoint{gp mark 7}{(2.492,2.392)}
\gppoint{gp mark 7}{(3.519,2.413)}
\gpcolor{color=gp lt color border}
\draw[gp path] (1.054,3.593)--(1.054,0.432)--(3.930,0.432)--(3.930,3.593)--cycle;
\gpdefrectangularnode{gp plot 1}{\pgfpoint{1.054cm}{0.432cm}}{\pgfpoint{3.930cm}{3.593cm}}
\end{tikzpicture}
}
					\subfloat[]{\begin{tikzpicture}[gnuplot]
\tikzset{every node/.append style={font={\fontsize{7.0pt}{8.4pt}\selectfont}}}
\path (0.000,0.000) rectangle (4.318,3.810);
\gpcolor{color=gp lt color axes}
\gpsetlinetype{gp lt axes}
\gpsetdashtype{gp dt axes}
\gpsetlinewidth{0.50}
\draw[gp path] (1.054,0.432)--(3.930,0.432);
\gpcolor{color=gp lt color border}
\gpsetlinetype{gp lt border}
\gpsetdashtype{gp dt solid}
\gpsetlinewidth{1.00}
\draw[gp path] (1.054,0.432)--(1.234,0.432);
\draw[gp path] (3.930,0.432)--(3.750,0.432);
\node[gp node right] at (0.925,0.432) {$0.5$};
\gpcolor{color=gp lt color axes}
\gpsetlinetype{gp lt axes}
\gpsetdashtype{gp dt axes}
\gpsetlinewidth{0.50}
\draw[gp path] (1.054,0.959)--(3.930,0.959);
\gpcolor{color=gp lt color border}
\gpsetlinetype{gp lt border}
\gpsetdashtype{gp dt solid}
\gpsetlinewidth{1.00}
\draw[gp path] (1.054,0.959)--(1.234,0.959);
\draw[gp path] (3.930,0.959)--(3.750,0.959);
\node[gp node right] at (0.925,0.959) {$0.55$};
\gpcolor{color=gp lt color axes}
\gpsetlinetype{gp lt axes}
\gpsetdashtype{gp dt axes}
\gpsetlinewidth{0.50}
\draw[gp path] (1.054,1.486)--(3.930,1.486);
\gpcolor{color=gp lt color border}
\gpsetlinetype{gp lt border}
\gpsetdashtype{gp dt solid}
\gpsetlinewidth{1.00}
\draw[gp path] (1.054,1.486)--(1.234,1.486);
\draw[gp path] (3.930,1.486)--(3.750,1.486);
\node[gp node right] at (0.925,1.486) {$0.6$};
\gpcolor{color=gp lt color axes}
\gpsetlinetype{gp lt axes}
\gpsetdashtype{gp dt axes}
\gpsetlinewidth{0.50}
\draw[gp path] (1.054,2.013)--(3.930,2.013);
\gpcolor{color=gp lt color border}
\gpsetlinetype{gp lt border}
\gpsetdashtype{gp dt solid}
\gpsetlinewidth{1.00}
\draw[gp path] (1.054,2.013)--(1.234,2.013);
\draw[gp path] (3.930,2.013)--(3.750,2.013);
\node[gp node right] at (0.925,2.013) {$0.65$};
\gpcolor{color=gp lt color axes}
\gpsetlinetype{gp lt axes}
\gpsetdashtype{gp dt axes}
\gpsetlinewidth{0.50}
\draw[gp path] (1.054,2.539)--(3.930,2.539);
\gpcolor{color=gp lt color border}
\gpsetlinetype{gp lt border}
\gpsetdashtype{gp dt solid}
\gpsetlinewidth{1.00}
\draw[gp path] (1.054,2.539)--(1.234,2.539);
\draw[gp path] (3.930,2.539)--(3.750,2.539);
\node[gp node right] at (0.925,2.539) {$0.7$};
\gpcolor{color=gp lt color axes}
\gpsetlinetype{gp lt axes}
\gpsetdashtype{gp dt axes}
\gpsetlinewidth{0.50}
\draw[gp path] (1.054,3.066)--(3.930,3.066);
\gpcolor{color=gp lt color border}
\gpsetlinetype{gp lt border}
\gpsetdashtype{gp dt solid}
\gpsetlinewidth{1.00}
\draw[gp path] (1.054,3.066)--(1.234,3.066);
\draw[gp path] (3.930,3.066)--(3.750,3.066);
\node[gp node right] at (0.925,3.066) {$0.75$};
\gpcolor{color=gp lt color axes}
\gpsetlinetype{gp lt axes}
\gpsetdashtype{gp dt axes}
\gpsetlinewidth{0.50}
\draw[gp path] (1.054,3.593)--(3.930,3.593);
\gpcolor{color=gp lt color border}
\gpsetlinetype{gp lt border}
\gpsetdashtype{gp dt solid}
\gpsetlinewidth{1.00}
\draw[gp path] (1.054,3.593)--(1.234,3.593);
\draw[gp path] (3.930,3.593)--(3.750,3.593);
\node[gp node right] at (0.925,3.593) {$0.8$};
\gpcolor{color=gp lt color axes}
\gpsetlinetype{gp lt axes}
\gpsetdashtype{gp dt axes}
\gpsetlinewidth{0.50}
\draw[gp path] (1.465,0.432)--(1.465,3.593);
\gpcolor{color=gp lt color border}
\gpsetlinetype{gp lt border}
\gpsetdashtype{gp dt solid}
\gpsetlinewidth{1.00}
\draw[gp path] (1.465,0.432)--(1.465,0.612);
\draw[gp path] (1.465,3.593)--(1.465,3.413);
\node[gp node center] at (1.465,0.216) {R1};
\gpcolor{color=gp lt color axes}
\gpsetlinetype{gp lt axes}
\gpsetdashtype{gp dt axes}
\gpsetlinewidth{0.50}
\draw[gp path] (2.492,0.432)--(2.492,3.593);
\gpcolor{color=gp lt color border}
\gpsetlinetype{gp lt border}
\gpsetdashtype{gp dt solid}
\gpsetlinewidth{1.00}
\draw[gp path] (2.492,0.432)--(2.492,0.612);
\draw[gp path] (2.492,3.593)--(2.492,3.413);
\node[gp node center] at (2.492,0.216) {R2};
\gpcolor{color=gp lt color axes}
\gpsetlinetype{gp lt axes}
\gpsetdashtype{gp dt axes}
\gpsetlinewidth{0.50}
\draw[gp path] (3.519,0.432)--(3.519,3.413)--(3.519,3.593);
\gpcolor{color=gp lt color border}
\gpsetlinetype{gp lt border}
\gpsetdashtype{gp dt solid}
\gpsetlinewidth{1.00}
\draw[gp path] (3.519,0.432)--(3.519,0.612);
\draw[gp path] (3.519,3.593)--(3.519,3.413);
\node[gp node center] at (3.519,0.216) {R3};
\draw[gp path] (1.054,3.593)--(1.054,0.432)--(3.930,0.432)--(3.930,3.593)--cycle;
\node[gp node center,rotate=-270] at (0.204,2.012) {Completion time [sec]};
\gpcolor{rgb color={0.580,0.000,0.827}}
\draw[gp path] (1.465,1.486)--(1.465,2.329);
\draw[gp path] (1.375,1.486)--(1.555,1.486);
\draw[gp path] (1.375,2.329)--(1.555,2.329);
\draw[gp path] (2.492,1.496)--(2.492,2.971);
\draw[gp path] (2.402,1.496)--(2.582,1.496);
\draw[gp path] (2.402,2.971)--(2.582,2.971);
\draw[gp path] (3.519,1.549)--(3.519,2.771);
\draw[gp path] (3.429,1.549)--(3.609,1.549);
\draw[gp path] (3.429,2.771)--(3.609,2.771);
\gpsetpointsize{4.00}
\gppoint{gp mark 1}{(1.465,1.907)}
\gppoint{gp mark 1}{(2.492,2.234)}
\gppoint{gp mark 1}{(3.519,2.160)}
\gpsetpointsize{2.40}
\gppoint{gp mark 7}{(1.465,1.907)}
\gppoint{gp mark 7}{(2.492,2.234)}
\gppoint{gp mark 7}{(3.519,2.160)}
\gpcolor{color=gp lt color border}
\draw[gp path] (1.054,3.593)--(1.054,0.432)--(3.930,0.432)--(3.930,3.593)--cycle;
\gpdefrectangularnode{gp plot 1}{\pgfpoint{1.054cm}{0.432cm}}{\pgfpoint{3.930cm}{3.593cm}}
\end{tikzpicture}
}
			\end{tabular}}
		\end{tabular}
		\caption{Completion time of fiber-path (a) establishment and (b) release.}
		\label{fig:results-path-control}
	\end{figure}

	We first executed the fiber-path establishment and release on each route $10$ times using our NBI. The operations were always successfully carried out by atomically and concurrently configuring multiple OCSes. The completion time of each operation is shown in Fig.~\ref{fig:results-path-control}. As shown in the figure, the completion time increases in the order of R1, R2, and R3. This is because the maximum time to configure each OCS on the route tends to become long as the number of OCSes to be configured increases. However, we found that the fiber paths were established or released on all routes within $1.0$ second under all settings. These results indicate that our controller efficiently operates fiber paths in a real OCS-based network environment and satisfies requirements~\ref{sec:req}.\ref{subsec:req-a} and~\ref{sec:req}.\ref{subsec:req-b}.
	
\begin{figure}[t]
		\centering
		\begin{tabular}{cc}
			{\begin{tabular}{@{}c@{}}
					\subfloat[]{\begin{tikzpicture}[gnuplot]
\tikzset{every node/.append style={font={\fontsize{7.0pt}{8.4pt}\selectfont}}}
\path (0.000,0.000) rectangle (4.318,3.810);
\gpcolor{color=gp lt color axes}
\gpsetlinetype{gp lt axes}
\gpsetdashtype{gp dt axes}
\gpsetlinewidth{0.50}
\draw[gp path] (0.925,0.432)--(3.930,0.432);
\gpcolor{color=gp lt color border}
\gpsetlinetype{gp lt border}
\gpsetdashtype{gp dt solid}
\gpsetlinewidth{1.00}
\draw[gp path] (0.925,0.432)--(1.105,0.432);
\draw[gp path] (3.930,0.432)--(3.750,0.432);
\node[gp node right] at (0.796,0.432) {$0.5$};
\gpcolor{color=gp lt color axes}
\gpsetlinetype{gp lt axes}
\gpsetdashtype{gp dt axes}
\gpsetlinewidth{0.50}
\draw[gp path] (0.925,1.064)--(3.930,1.064);
\gpcolor{color=gp lt color border}
\gpsetlinetype{gp lt border}
\gpsetdashtype{gp dt solid}
\gpsetlinewidth{1.00}
\draw[gp path] (0.925,1.064)--(1.105,1.064);
\draw[gp path] (3.930,1.064)--(3.750,1.064);
\node[gp node right] at (0.796,1.064) {$0.6$};
\gpcolor{color=gp lt color axes}
\gpsetlinetype{gp lt axes}
\gpsetdashtype{gp dt axes}
\gpsetlinewidth{0.50}
\draw[gp path] (0.925,1.696)--(3.930,1.696);
\gpcolor{color=gp lt color border}
\gpsetlinetype{gp lt border}
\gpsetdashtype{gp dt solid}
\gpsetlinewidth{1.00}
\draw[gp path] (0.925,1.696)--(1.105,1.696);
\draw[gp path] (3.930,1.696)--(3.750,1.696);
\node[gp node right] at (0.796,1.696) {$0.7$};
\gpcolor{color=gp lt color axes}
\gpsetlinetype{gp lt axes}
\gpsetdashtype{gp dt axes}
\gpsetlinewidth{0.50}
\draw[gp path] (0.925,2.329)--(3.930,2.329);
\gpcolor{color=gp lt color border}
\gpsetlinetype{gp lt border}
\gpsetdashtype{gp dt solid}
\gpsetlinewidth{1.00}
\draw[gp path] (0.925,2.329)--(1.105,2.329);
\draw[gp path] (3.930,2.329)--(3.750,2.329);
\node[gp node right] at (0.796,2.329) {$0.8$};
\gpcolor{color=gp lt color axes}
\gpsetlinetype{gp lt axes}
\gpsetdashtype{gp dt axes}
\gpsetlinewidth{0.50}
\draw[gp path] (0.925,2.961)--(3.930,2.961);
\gpcolor{color=gp lt color border}
\gpsetlinetype{gp lt border}
\gpsetdashtype{gp dt solid}
\gpsetlinewidth{1.00}
\draw[gp path] (0.925,2.961)--(1.105,2.961);
\draw[gp path] (3.930,2.961)--(3.750,2.961);
\node[gp node right] at (0.796,2.961) {$0.9$};
\gpcolor{color=gp lt color axes}
\gpsetlinetype{gp lt axes}
\gpsetdashtype{gp dt axes}
\gpsetlinewidth{0.50}
\draw[gp path] (0.925,3.593)--(3.930,3.593);
\gpcolor{color=gp lt color border}
\gpsetlinetype{gp lt border}
\gpsetdashtype{gp dt solid}
\gpsetlinewidth{1.00}
\draw[gp path] (0.925,3.593)--(1.105,3.593);
\draw[gp path] (3.930,3.593)--(3.750,3.593);
\node[gp node right] at (0.796,3.593) {$1$};
\gpcolor{color=gp lt color axes}
\gpsetlinetype{gp lt axes}
\gpsetdashtype{gp dt axes}
\gpsetlinewidth{0.50}
\draw[gp path] (1.354,0.432)--(1.354,3.593);
\gpcolor{color=gp lt color border}
\gpsetlinetype{gp lt border}
\gpsetdashtype{gp dt solid}
\gpsetlinewidth{1.00}
\draw[gp path] (1.354,0.432)--(1.354,0.612);
\draw[gp path] (1.354,3.593)--(1.354,3.413);
\node[gp node center] at (1.354,0.216) {M=1};
\gpcolor{color=gp lt color axes}
\gpsetlinetype{gp lt axes}
\gpsetdashtype{gp dt axes}
\gpsetlinewidth{0.50}
\draw[gp path] (2.428,0.432)--(2.428,3.593);
\gpcolor{color=gp lt color border}
\gpsetlinetype{gp lt border}
\gpsetdashtype{gp dt solid}
\gpsetlinewidth{1.00}
\draw[gp path] (2.428,0.432)--(2.428,0.612);
\draw[gp path] (2.428,3.593)--(2.428,3.413);
\node[gp node center] at (2.428,0.216) {M=2};
\gpcolor{color=gp lt color axes}
\gpsetlinetype{gp lt axes}
\gpsetdashtype{gp dt axes}
\gpsetlinewidth{0.50}
\draw[gp path] (3.501,0.432)--(3.501,3.413)--(3.501,3.593);
\gpcolor{color=gp lt color border}
\gpsetlinetype{gp lt border}
\gpsetdashtype{gp dt solid}
\gpsetlinewidth{1.00}
\draw[gp path] (3.501,0.432)--(3.501,0.612);
\draw[gp path] (3.501,3.593)--(3.501,3.413);
\node[gp node center] at (3.501,0.216) {M=3};
\draw[gp path] (0.925,3.593)--(0.925,0.432)--(3.930,0.432)--(3.930,3.593)--cycle;
\node[gp node center,rotate=-270] at (0.204,2.012) {Completion time [sec]};
\gpcolor{rgb color={0.580,0.000,0.827}}
\draw[gp path] (1.354,1.663)--(1.354,2.434);
\draw[gp path] (1.264,1.663)--(1.444,1.663);
\draw[gp path] (1.264,2.434)--(1.444,2.434);
\draw[gp path] (2.428,0.956)--(2.428,2.195);
\draw[gp path] (2.338,0.956)--(2.518,0.956);
\draw[gp path] (2.338,2.195)--(2.518,2.195);
\draw[gp path] (3.501,1.337)--(3.501,2.753);
\draw[gp path] (3.411,1.337)--(3.591,1.337);
\draw[gp path] (3.411,2.753)--(3.591,2.753);
\gpsetpointsize{4.00}
\gppoint{gp mark 1}{(1.354,2.049)}
\gppoint{gp mark 1}{(2.428,1.576)}
\gppoint{gp mark 1}{(3.501,2.045)}
\gpsetpointsize{2.40}
\gppoint{gp mark 7}{(1.354,2.049)}
\gppoint{gp mark 7}{(2.428,1.576)}
\gppoint{gp mark 7}{(3.501,2.045)}
\gpcolor{color=gp lt color border}
\draw[gp path] (0.925,3.593)--(0.925,0.432)--(3.930,0.432)--(3.930,3.593)--cycle;
\gpdefrectangularnode{gp plot 1}{\pgfpoint{0.925cm}{0.432cm}}{\pgfpoint{3.930cm}{3.593cm}}
\end{tikzpicture}
}
					\subfloat[]{\begin{tikzpicture}[gnuplot]
\tikzset{every node/.append style={font={\fontsize{7.0pt}{8.4pt}\selectfont}}}
\path (0.000,0.000) rectangle (4.318,3.810);
\gpcolor{color=gp lt color axes}
\gpsetlinetype{gp lt axes}
\gpsetdashtype{gp dt axes}
\gpsetlinewidth{0.50}
\draw[gp path] (0.925,0.432)--(3.930,0.432);
\gpcolor{color=gp lt color border}
\gpsetlinetype{gp lt border}
\gpsetdashtype{gp dt solid}
\gpsetlinewidth{1.00}
\draw[gp path] (0.925,0.432)--(1.105,0.432);
\draw[gp path] (3.930,0.432)--(3.750,0.432);
\node[gp node right] at (0.796,0.432) {$0.2$};
\gpcolor{color=gp lt color axes}
\gpsetlinetype{gp lt axes}
\gpsetdashtype{gp dt axes}
\gpsetlinewidth{0.50}
\draw[gp path] (0.925,0.959)--(3.930,0.959);
\gpcolor{color=gp lt color border}
\gpsetlinetype{gp lt border}
\gpsetdashtype{gp dt solid}
\gpsetlinewidth{1.00}
\draw[gp path] (0.925,0.959)--(1.105,0.959);
\draw[gp path] (3.930,0.959)--(3.750,0.959);
\node[gp node right] at (0.796,0.959) {$0.3$};
\gpcolor{color=gp lt color axes}
\gpsetlinetype{gp lt axes}
\gpsetdashtype{gp dt axes}
\gpsetlinewidth{0.50}
\draw[gp path] (0.925,1.486)--(3.930,1.486);
\gpcolor{color=gp lt color border}
\gpsetlinetype{gp lt border}
\gpsetdashtype{gp dt solid}
\gpsetlinewidth{1.00}
\draw[gp path] (0.925,1.486)--(1.105,1.486);
\draw[gp path] (3.930,1.486)--(3.750,1.486);
\node[gp node right] at (0.796,1.486) {$0.4$};
\gpcolor{color=gp lt color axes}
\gpsetlinetype{gp lt axes}
\gpsetdashtype{gp dt axes}
\gpsetlinewidth{0.50}
\draw[gp path] (0.925,2.012)--(3.930,2.012);
\gpcolor{color=gp lt color border}
\gpsetlinetype{gp lt border}
\gpsetdashtype{gp dt solid}
\gpsetlinewidth{1.00}
\draw[gp path] (0.925,2.012)--(1.105,2.012);
\draw[gp path] (3.930,2.012)--(3.750,2.012);
\node[gp node right] at (0.796,2.012) {$0.5$};
\gpcolor{color=gp lt color axes}
\gpsetlinetype{gp lt axes}
\gpsetdashtype{gp dt axes}
\gpsetlinewidth{0.50}
\draw[gp path] (0.925,2.539)--(3.930,2.539);
\gpcolor{color=gp lt color border}
\gpsetlinetype{gp lt border}
\gpsetdashtype{gp dt solid}
\gpsetlinewidth{1.00}
\draw[gp path] (0.925,2.539)--(1.105,2.539);
\draw[gp path] (3.930,2.539)--(3.750,2.539);
\node[gp node right] at (0.796,2.539) {$0.6$};
\gpcolor{color=gp lt color axes}
\gpsetlinetype{gp lt axes}
\gpsetdashtype{gp dt axes}
\gpsetlinewidth{0.50}
\draw[gp path] (0.925,3.066)--(3.930,3.066);
\gpcolor{color=gp lt color border}
\gpsetlinetype{gp lt border}
\gpsetdashtype{gp dt solid}
\gpsetlinewidth{1.00}
\draw[gp path] (0.925,3.066)--(1.105,3.066);
\draw[gp path] (3.930,3.066)--(3.750,3.066);
\node[gp node right] at (0.796,3.066) {$0.7$};
\gpcolor{color=gp lt color axes}
\gpsetlinetype{gp lt axes}
\gpsetdashtype{gp dt axes}
\gpsetlinewidth{0.50}
\draw[gp path] (0.925,3.593)--(3.930,3.593);
\gpcolor{color=gp lt color border}
\gpsetlinetype{gp lt border}
\gpsetdashtype{gp dt solid}
\gpsetlinewidth{1.00}
\draw[gp path] (0.925,3.593)--(1.105,3.593);
\draw[gp path] (3.930,3.593)--(3.750,3.593);
\node[gp node right] at (0.796,3.593) {$0.8$};
\gpcolor{color=gp lt color axes}
\gpsetlinetype{gp lt axes}
\gpsetdashtype{gp dt axes}
\gpsetlinewidth{0.50}
\draw[gp path] (1.354,0.432)--(1.354,3.593);
\gpcolor{color=gp lt color border}
\gpsetlinetype{gp lt border}
\gpsetdashtype{gp dt solid}
\gpsetlinewidth{1.00}
\draw[gp path] (1.354,0.432)--(1.354,0.612);
\draw[gp path] (1.354,3.593)--(1.354,3.413);
\node[gp node center] at (1.354,0.216) {M=1};
\gpcolor{color=gp lt color axes}
\gpsetlinetype{gp lt axes}
\gpsetdashtype{gp dt axes}
\gpsetlinewidth{0.50}
\draw[gp path] (2.428,0.432)--(2.428,3.593);
\gpcolor{color=gp lt color border}
\gpsetlinetype{gp lt border}
\gpsetdashtype{gp dt solid}
\gpsetlinewidth{1.00}
\draw[gp path] (2.428,0.432)--(2.428,0.612);
\draw[gp path] (2.428,3.593)--(2.428,3.413);
\node[gp node center] at (2.428,0.216) {M=2};
\gpcolor{color=gp lt color axes}
\gpsetlinetype{gp lt axes}
\gpsetdashtype{gp dt axes}
\gpsetlinewidth{0.50}
\draw[gp path] (3.501,0.432)--(3.501,3.413)--(3.501,3.593);
\gpcolor{color=gp lt color border}
\gpsetlinetype{gp lt border}
\gpsetdashtype{gp dt solid}
\gpsetlinewidth{1.00}
\draw[gp path] (3.501,0.432)--(3.501,0.612);
\draw[gp path] (3.501,3.593)--(3.501,3.413);
\node[gp node center] at (3.501,0.216) {M=3};
\draw[gp path] (0.925,3.593)--(0.925,0.432)--(3.930,0.432)--(3.930,3.593)--cycle;
\node[gp node center,rotate=-270] at (0.204,2.012) {Completion time [sec]};
\gpcolor{rgb color={0.580,0.000,0.827}}
\draw[gp path] (1.354,2.055)--(1.354,3.045);
\draw[gp path] (1.264,2.055)--(1.444,2.055);
\draw[gp path] (1.264,3.045)--(1.444,3.045);
\draw[gp path] (2.428,2.471)--(2.428,3.314);
\draw[gp path] (2.338,2.471)--(2.518,2.471);
\draw[gp path] (2.338,3.314)--(2.518,3.314);
\draw[gp path] (3.501,1.470)--(3.501,2.260);
\draw[gp path] (3.411,1.470)--(3.591,1.470);
\draw[gp path] (3.411,2.260)--(3.591,2.260);
\gpsetpointsize{4.00}
\gppoint{gp mark 1}{(1.354,2.550)}
\gppoint{gp mark 1}{(2.428,2.892)}
\gppoint{gp mark 1}{(3.501,1.865)}
\gpsetpointsize{2.40}
\gppoint{gp mark 7}{(1.354,2.550)}
\gppoint{gp mark 7}{(2.428,2.892)}
\gppoint{gp mark 7}{(3.501,1.865)}
\gpcolor{color=gp lt color border}
\draw[gp path] (0.925,3.593)--(0.925,0.432)--(3.930,0.432)--(3.930,3.593)--cycle;
\gpdefrectangularnode{gp plot 1}{\pgfpoint{0.925cm}{0.432cm}}{\pgfpoint{3.930cm}{3.593cm}}
\end{tikzpicture}
}
			\end{tabular}}
		\end{tabular}
		\caption{Completion time of roll-back operation during fiber-path (a) establishment and (b) release.}
		\label{fig:results-rollback}
	\end{figure}
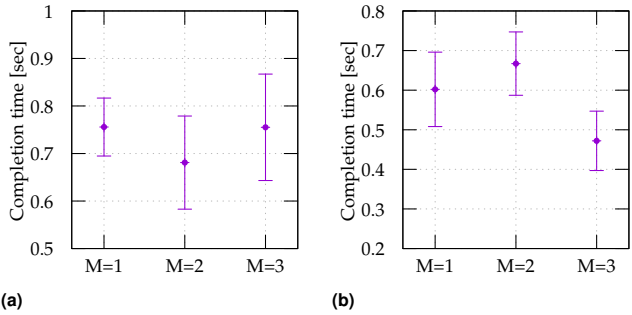

	Second, we checked if roll-back operations could be successfully carried out by our controller when OCS operation failed. The evaluation was conducted on R3. Before establishing and releasing a fiber path on R3, we intentionally stopped the NETCONF server on $M$ OCSes to simulate the OCS failure. In this case, the controller should receive the timeout error from $M$ OCSes during the path operations and immediately execute roll-back operations by reverting to the configurations of normal OCSes. As we expected, we observed that roll-back operations could be successfully executed by our controller in all cases. The time required for the roll-back operations under different $M$ is shown in Fig.~\ref{fig:results-rollback}. As shown in the figure, roll-back operations always completed within $0.90$ seconds under all settings. We could achieve this by concurrently configuring OCSes from our controller. These results indicate that our controller always consistently and efficiently manages the OCS-based network and satisfies requirement~\ref{sec:req}.\ref{subsec:req-c}.
	
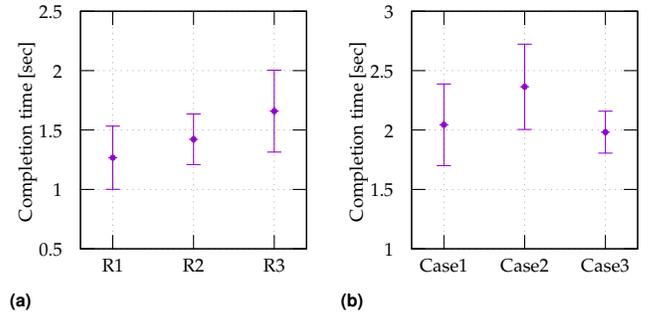
\begin{figure}[t]
		\centering
		\begin{tabular}{cc}
			{\begin{tabular}{@{}c@{}}
					\subfloat[]{\begin{tikzpicture}[gnuplot]
\tikzset{every node/.append style={font={\fontsize{7.0pt}{8.4pt}\selectfont}}}
\path (0.000,0.000) rectangle (4.318,3.810);
\gpcolor{color=gp lt color axes}
\gpsetlinetype{gp lt axes}
\gpsetdashtype{gp dt axes}
\gpsetlinewidth{0.50}
\draw[gp path] (0.925,0.432)--(3.930,0.432);
\gpcolor{color=gp lt color border}
\gpsetlinetype{gp lt border}
\gpsetdashtype{gp dt solid}
\gpsetlinewidth{1.00}
\draw[gp path] (0.925,0.432)--(1.105,0.432);
\draw[gp path] (3.930,0.432)--(3.750,0.432);
\node[gp node right] at (0.796,0.432) {$0.5$};
\gpcolor{color=gp lt color axes}
\gpsetlinetype{gp lt axes}
\gpsetdashtype{gp dt axes}
\gpsetlinewidth{0.50}
\draw[gp path] (0.925,1.222)--(3.930,1.222);
\gpcolor{color=gp lt color border}
\gpsetlinetype{gp lt border}
\gpsetdashtype{gp dt solid}
\gpsetlinewidth{1.00}
\draw[gp path] (0.925,1.222)--(1.105,1.222);
\draw[gp path] (3.930,1.222)--(3.750,1.222);
\node[gp node right] at (0.796,1.222) {$1$};
\gpcolor{color=gp lt color axes}
\gpsetlinetype{gp lt axes}
\gpsetdashtype{gp dt axes}
\gpsetlinewidth{0.50}
\draw[gp path] (0.925,2.013)--(3.930,2.013);
\gpcolor{color=gp lt color border}
\gpsetlinetype{gp lt border}
\gpsetdashtype{gp dt solid}
\gpsetlinewidth{1.00}
\draw[gp path] (0.925,2.013)--(1.105,2.013);
\draw[gp path] (3.930,2.013)--(3.750,2.013);
\node[gp node right] at (0.796,2.013) {$1.5$};
\gpcolor{color=gp lt color axes}
\gpsetlinetype{gp lt axes}
\gpsetdashtype{gp dt axes}
\gpsetlinewidth{0.50}
\draw[gp path] (0.925,2.803)--(3.930,2.803);
\gpcolor{color=gp lt color border}
\gpsetlinetype{gp lt border}
\gpsetdashtype{gp dt solid}
\gpsetlinewidth{1.00}
\draw[gp path] (0.925,2.803)--(1.105,2.803);
\draw[gp path] (3.930,2.803)--(3.750,2.803);
\node[gp node right] at (0.796,2.803) {$2$};
\gpcolor{color=gp lt color axes}
\gpsetlinetype{gp lt axes}
\gpsetdashtype{gp dt axes}
\gpsetlinewidth{0.50}
\draw[gp path] (0.925,3.593)--(3.930,3.593);
\gpcolor{color=gp lt color border}
\gpsetlinetype{gp lt border}
\gpsetdashtype{gp dt solid}
\gpsetlinewidth{1.00}
\draw[gp path] (0.925,3.593)--(1.105,3.593);
\draw[gp path] (3.930,3.593)--(3.750,3.593);
\node[gp node right] at (0.796,3.593) {$2.5$};
\gpcolor{color=gp lt color axes}
\gpsetlinetype{gp lt axes}
\gpsetdashtype{gp dt axes}
\gpsetlinewidth{0.50}
\draw[gp path] (1.354,0.432)--(1.354,3.593);
\gpcolor{color=gp lt color border}
\gpsetlinetype{gp lt border}
\gpsetdashtype{gp dt solid}
\gpsetlinewidth{1.00}
\draw[gp path] (1.354,0.432)--(1.354,0.612);
\draw[gp path] (1.354,3.593)--(1.354,3.413);
\node[gp node center] at (1.354,0.216) {R1};
\gpcolor{color=gp lt color axes}
\gpsetlinetype{gp lt axes}
\gpsetdashtype{gp dt axes}
\gpsetlinewidth{0.50}
\draw[gp path] (2.428,0.432)--(2.428,3.593);
\gpcolor{color=gp lt color border}
\gpsetlinetype{gp lt border}
\gpsetdashtype{gp dt solid}
\gpsetlinewidth{1.00}
\draw[gp path] (2.428,0.432)--(2.428,0.612);
\draw[gp path] (2.428,3.593)--(2.428,3.413);
\node[gp node center] at (2.428,0.216) {R2};
\gpcolor{color=gp lt color axes}
\gpsetlinetype{gp lt axes}
\gpsetdashtype{gp dt axes}
\gpsetlinewidth{0.50}
\draw[gp path] (3.501,0.432)--(3.501,3.413)--(3.501,3.593);
\gpcolor{color=gp lt color border}
\gpsetlinetype{gp lt border}
\gpsetdashtype{gp dt solid}
\gpsetlinewidth{1.00}
\draw[gp path] (3.501,0.432)--(3.501,0.612);
\draw[gp path] (3.501,3.593)--(3.501,3.413);
\node[gp node center] at (3.501,0.216) {R3};
\draw[gp path] (0.925,3.593)--(0.925,0.432)--(3.930,0.432)--(3.930,3.593)--cycle;
\node[gp node center,rotate=-270] at (0.204,2.012) {Completion time [sec]};
\gpcolor{rgb color={0.580,0.000,0.827}}
\draw[gp path] (1.354,1.224)--(1.354,2.065);
\draw[gp path] (1.264,1.224)--(1.444,1.224);
\draw[gp path] (1.264,2.065)--(1.444,2.065);
\draw[gp path] (2.428,1.553)--(2.428,2.226);
\draw[gp path] (2.338,1.553)--(2.518,1.553);
\draw[gp path] (2.338,2.226)--(2.518,2.226);
\draw[gp path] (3.501,1.719)--(3.501,2.809);
\draw[gp path] (3.411,1.719)--(3.591,1.719);
\draw[gp path] (3.411,2.809)--(3.591,2.809);
\gpsetpointsize{4.00}
\gppoint{gp mark 1}{(1.354,1.644)}
\gppoint{gp mark 1}{(2.428,1.889)}
\gppoint{gp mark 1}{(3.501,2.264)}
\gpsetpointsize{2.40}
\gppoint{gp mark 7}{(1.354,1.644)}
\gppoint{gp mark 7}{(2.428,1.889)}
\gppoint{gp mark 7}{(3.501,2.264)}
\gpcolor{color=gp lt color border}
\draw[gp path] (0.925,3.593)--(0.925,0.432)--(3.930,0.432)--(3.930,3.593)--cycle;
\gpdefrectangularnode{gp plot 1}{\pgfpoint{0.925cm}{0.432cm}}{\pgfpoint{3.930cm}{3.593cm}}
\end{tikzpicture}
}
					\subfloat[]{\begin{tikzpicture}[gnuplot]
\tikzset{every node/.append style={font={\fontsize{7.0pt}{8.4pt}\selectfont}}}
\path (0.000,0.000) rectangle (4.318,3.810);
\gpcolor{color=gp lt color axes}
\gpsetlinetype{gp lt axes}
\gpsetdashtype{gp dt axes}
\gpsetlinewidth{0.50}
\draw[gp path] (0.925,0.432)--(3.930,0.432);
\gpcolor{color=gp lt color border}
\gpsetlinetype{gp lt border}
\gpsetdashtype{gp dt solid}
\gpsetlinewidth{1.00}
\draw[gp path] (0.925,0.432)--(1.105,0.432);
\draw[gp path] (3.930,0.432)--(3.750,0.432);
\node[gp node right] at (0.796,0.432) {$1$};
\gpcolor{color=gp lt color axes}
\gpsetlinetype{gp lt axes}
\gpsetdashtype{gp dt axes}
\gpsetlinewidth{0.50}
\draw[gp path] (0.925,1.222)--(3.930,1.222);
\gpcolor{color=gp lt color border}
\gpsetlinetype{gp lt border}
\gpsetdashtype{gp dt solid}
\gpsetlinewidth{1.00}
\draw[gp path] (0.925,1.222)--(1.105,1.222);
\draw[gp path] (3.930,1.222)--(3.750,1.222);
\node[gp node right] at (0.796,1.222) {$1.5$};
\gpcolor{color=gp lt color axes}
\gpsetlinetype{gp lt axes}
\gpsetdashtype{gp dt axes}
\gpsetlinewidth{0.50}
\draw[gp path] (0.925,2.013)--(3.930,2.013);
\gpcolor{color=gp lt color border}
\gpsetlinetype{gp lt border}
\gpsetdashtype{gp dt solid}
\gpsetlinewidth{1.00}
\draw[gp path] (0.925,2.013)--(1.105,2.013);
\draw[gp path] (3.930,2.013)--(3.750,2.013);
\node[gp node right] at (0.796,2.013) {$2$};
\gpcolor{color=gp lt color axes}
\gpsetlinetype{gp lt axes}
\gpsetdashtype{gp dt axes}
\gpsetlinewidth{0.50}
\draw[gp path] (0.925,2.803)--(3.930,2.803);
\gpcolor{color=gp lt color border}
\gpsetlinetype{gp lt border}
\gpsetdashtype{gp dt solid}
\gpsetlinewidth{1.00}
\draw[gp path] (0.925,2.803)--(1.105,2.803);
\draw[gp path] (3.930,2.803)--(3.750,2.803);
\node[gp node right] at (0.796,2.803) {$2.5$};
\gpcolor{color=gp lt color axes}
\gpsetlinetype{gp lt axes}
\gpsetdashtype{gp dt axes}
\gpsetlinewidth{0.50}
\draw[gp path] (0.925,3.593)--(3.930,3.593);
\gpcolor{color=gp lt color border}
\gpsetlinetype{gp lt border}
\gpsetdashtype{gp dt solid}
\gpsetlinewidth{1.00}
\draw[gp path] (0.925,3.593)--(1.105,3.593);
\draw[gp path] (3.930,3.593)--(3.750,3.593);
\node[gp node right] at (0.796,3.593) {$3$};
\gpcolor{color=gp lt color axes}
\gpsetlinetype{gp lt axes}
\gpsetdashtype{gp dt axes}
\gpsetlinewidth{0.50}
\draw[gp path] (1.354,0.432)--(1.354,3.593);
\gpcolor{color=gp lt color border}
\gpsetlinetype{gp lt border}
\gpsetdashtype{gp dt solid}
\gpsetlinewidth{1.00}
\draw[gp path] (1.354,0.432)--(1.354,0.612);
\draw[gp path] (1.354,3.593)--(1.354,3.413);
\node[gp node center] at (1.354,0.216) {Case1};
\gpcolor{color=gp lt color axes}
\gpsetlinetype{gp lt axes}
\gpsetdashtype{gp dt axes}
\gpsetlinewidth{0.50}
\draw[gp path] (2.428,0.432)--(2.428,3.593);
\gpcolor{color=gp lt color border}
\gpsetlinetype{gp lt border}
\gpsetdashtype{gp dt solid}
\gpsetlinewidth{1.00}
\draw[gp path] (2.428,0.432)--(2.428,0.612);
\draw[gp path] (2.428,3.593)--(2.428,3.413);
\node[gp node center] at (2.428,0.216) {Case2};
\gpcolor{color=gp lt color axes}
\gpsetlinetype{gp lt axes}
\gpsetdashtype{gp dt axes}
\gpsetlinewidth{0.50}
\draw[gp path] (3.501,0.432)--(3.501,3.413)--(3.501,3.593);
\gpcolor{color=gp lt color border}
\gpsetlinetype{gp lt border}
\gpsetdashtype{gp dt solid}
\gpsetlinewidth{1.00}
\draw[gp path] (3.501,0.432)--(3.501,0.612);
\draw[gp path] (3.501,3.593)--(3.501,3.413);
\node[gp node center] at (3.501,0.216) {Case3};
\draw[gp path] (0.925,3.593)--(0.925,0.432)--(3.930,0.432)--(3.930,3.593)--cycle;
\node[gp node center,rotate=-270] at (0.204,2.012) {Completion time [sec]};
\gpcolor{rgb color={0.580,0.000,0.827}}
\draw[gp path] (1.354,1.539)--(1.354,2.625);
\draw[gp path] (1.264,1.539)--(1.444,1.539);
\draw[gp path] (1.264,2.625)--(1.444,2.625);
\draw[gp path] (2.428,2.020)--(2.428,3.155);
\draw[gp path] (2.338,2.020)--(2.518,2.020);
\draw[gp path] (2.338,3.155)--(2.518,3.155);
\draw[gp path] (3.501,1.705)--(3.501,2.263);
\draw[gp path] (3.411,1.705)--(3.591,1.705);
\draw[gp path] (3.411,2.263)--(3.591,2.263);
\gpsetpointsize{4.00}
\gppoint{gp mark 1}{(1.354,2.082)}
\gppoint{gp mark 1}{(2.428,2.588)}
\gppoint{gp mark 1}{(3.501,1.984)}
\gpsetpointsize{2.40}
\gppoint{gp mark 7}{(1.354,2.082)}
\gppoint{gp mark 7}{(2.428,2.588)}
\gppoint{gp mark 7}{(3.501,1.984)}
\gpcolor{color=gp lt color border}
\draw[gp path] (0.925,3.593)--(0.925,0.432)--(3.930,0.432)--(3.930,3.593)--cycle;
\gpdefrectangularnode{gp plot 1}{\pgfpoint{0.925cm}{0.432cm}}{\pgfpoint{3.930cm}{3.593cm}}
\end{tikzpicture}
}
			\end{tabular}}
		\end{tabular}
		\caption{Completion time of automatic fiber-path (a) establishment and (b) restoration.}
		\label{fig:results-event}
	\end{figure}

	Next, we checked if automatic fiber-path control upon signal detection could be executed by our controller. First, we tried to automatically establish the fiber paths on R1, R2, and R3 upon signal detection. Specifically, once the signal higher than $-1.0$ dBm was observed at Rx port on OCS \#1 connected with terminal A, one of the above fiber paths is automatically established. We carried out this operation $10$ times by turning on the laser source of terminal A that transmitted a $5.9$ dBm signal. As a result, fiber paths were always successfully established upon signal detection at OCS \#1 by our controller. The time required to perform this operation is shown in Fig.~\ref{fig:results-event}(a). It includes the time to detect the signal, receive the notification from the OCS \#1, and complete the fiber-path establishment. As shown in the figure, fiber-path establishment completed within $2.0$ seconds under all settings. This result indicates the possibility of efficient operations on the basis of our defined MV-OCS controller to support use cases discussed in~\cite{igf, kani2025disaggregation}, that is, automatically accommodating user terminals on metro-access converged networks. From the above results, we found that our controller satisfies requirement~\ref{sec:req}.\ref{subsec:req-d}.
	
	Finally, we checked if restoration could automatically be executed upon signal degradation by our controller. Specifically, we tried to automatically restore fiber paths when OCS \#3 observed the signal degradation at its Rx ports connected with each route: R1, R2, and R3. We tried three cases: \textbf{Case 1}, \textbf{Case 2}, and \textbf{Case 3} are the restorations from R1 to R2, R2 to R3, and R3 to R1, respectively. Once the signal on each route became less than $-10.0$ dBm, one of the above three cases was carried out. We carried out this operation $10$ times in each case by plugging out a fiber connected with OCS \#3. The experimental results showed that automatic restoration was always successfully carried out by our controller. Though OCSes \#1 and \#3 are from different vendors, these events were successfully handled by our controller, which demonstrates the validity of our controller framework. The time required to perform the restoration in each case is shown in Fig.~\ref{fig:results-event}(b); it was always less than $3.0$ seconds. This result indicates that zero-touch restoration is made possible by our MV-OCS controller, which could reduce OPEX. From the above results, we found that our controller satisfies requirement~\ref{sec:req}.\ref{subsec:req-d}. To further accelerate the restoration time (e.g., to a few milliseconds), communication protocol should be improved for both vendor-proprietary APIs and our defined SBI. For solving this issue, adopting an EtherCAT technology could be one of the candidate solutions as proposed in~\cite{takano2024fast} and left for future work.
	
	\begin{figure}[t]
		\centering
		\includegraphics[width=.48\textwidth]{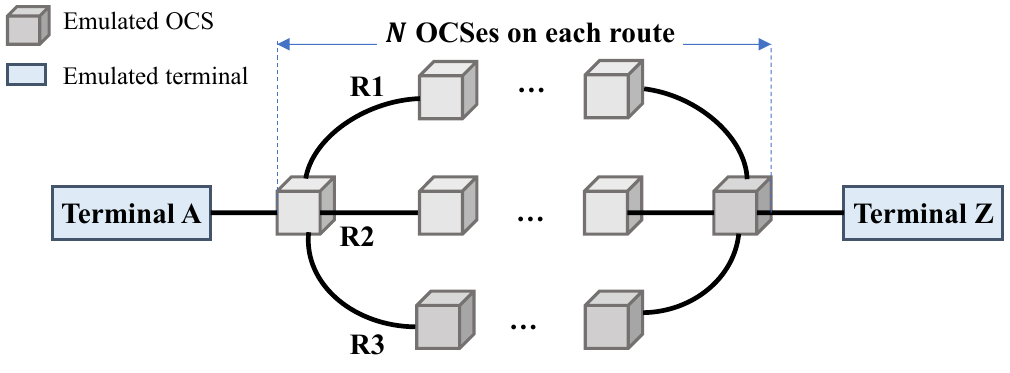}
		\caption{Large-scale emulated network.}
		\label{fig:exp-emu}
	\end{figure}

	\begin{figure}[t]
		\centering
		\begin{tabular}{cc}
			{\begin{tabular}{@{}c@{}}
					\subfloat[]{\begin{tikzpicture}[gnuplot]
\tikzset{every node/.append style={font={\fontsize{7.0pt}{8.4pt}\selectfont}}}
\path (0.000,0.000) rectangle (4.318,3.810);
\gpcolor{color=gp lt color axes}
\gpsetlinetype{gp lt axes}
\gpsetdashtype{gp dt axes}
\gpsetlinewidth{0.50}
\draw[gp path] (1.054,0.432)--(3.930,0.432);
\gpcolor{color=gp lt color border}
\gpsetlinetype{gp lt border}
\gpsetdashtype{gp dt solid}
\gpsetlinewidth{1.00}
\draw[gp path] (1.054,0.432)--(1.234,0.432);
\draw[gp path] (3.930,0.432)--(3.750,0.432);
\node[gp node right] at (0.925,0.432) {$0.7$};
\gpcolor{color=gp lt color axes}
\gpsetlinetype{gp lt axes}
\gpsetdashtype{gp dt axes}
\gpsetlinewidth{0.50}
\draw[gp path] (1.054,0.959)--(3.930,0.959);
\gpcolor{color=gp lt color border}
\gpsetlinetype{gp lt border}
\gpsetdashtype{gp dt solid}
\gpsetlinewidth{1.00}
\draw[gp path] (1.054,0.959)--(1.234,0.959);
\draw[gp path] (3.930,0.959)--(3.750,0.959);
\node[gp node right] at (0.925,0.959) {$0.75$};
\gpcolor{color=gp lt color axes}
\gpsetlinetype{gp lt axes}
\gpsetdashtype{gp dt axes}
\gpsetlinewidth{0.50}
\draw[gp path] (1.054,1.486)--(3.930,1.486);
\gpcolor{color=gp lt color border}
\gpsetlinetype{gp lt border}
\gpsetdashtype{gp dt solid}
\gpsetlinewidth{1.00}
\draw[gp path] (1.054,1.486)--(1.234,1.486);
\draw[gp path] (3.930,1.486)--(3.750,1.486);
\node[gp node right] at (0.925,1.486) {$0.8$};
\gpcolor{color=gp lt color axes}
\gpsetlinetype{gp lt axes}
\gpsetdashtype{gp dt axes}
\gpsetlinewidth{0.50}
\draw[gp path] (1.054,2.013)--(3.930,2.013);
\gpcolor{color=gp lt color border}
\gpsetlinetype{gp lt border}
\gpsetdashtype{gp dt solid}
\gpsetlinewidth{1.00}
\draw[gp path] (1.054,2.013)--(1.234,2.013);
\draw[gp path] (3.930,2.013)--(3.750,2.013);
\node[gp node right] at (0.925,2.013) {$0.85$};
\gpcolor{color=gp lt color axes}
\gpsetlinetype{gp lt axes}
\gpsetdashtype{gp dt axes}
\gpsetlinewidth{0.50}
\draw[gp path] (1.054,2.539)--(3.930,2.539);
\gpcolor{color=gp lt color border}
\gpsetlinetype{gp lt border}
\gpsetdashtype{gp dt solid}
\gpsetlinewidth{1.00}
\draw[gp path] (1.054,2.539)--(1.234,2.539);
\draw[gp path] (3.930,2.539)--(3.750,2.539);
\node[gp node right] at (0.925,2.539) {$0.9$};
\gpcolor{color=gp lt color axes}
\gpsetlinetype{gp lt axes}
\gpsetdashtype{gp dt axes}
\gpsetlinewidth{0.50}
\draw[gp path] (1.054,3.066)--(3.930,3.066);
\gpcolor{color=gp lt color border}
\gpsetlinetype{gp lt border}
\gpsetdashtype{gp dt solid}
\gpsetlinewidth{1.00}
\draw[gp path] (1.054,3.066)--(1.234,3.066);
\draw[gp path] (3.930,3.066)--(3.750,3.066);
\node[gp node right] at (0.925,3.066) {$0.95$};
\gpcolor{color=gp lt color axes}
\gpsetlinetype{gp lt axes}
\gpsetdashtype{gp dt axes}
\gpsetlinewidth{0.50}
\draw[gp path] (1.054,3.593)--(3.930,3.593);
\gpcolor{color=gp lt color border}
\gpsetlinetype{gp lt border}
\gpsetdashtype{gp dt solid}
\gpsetlinewidth{1.00}
\draw[gp path] (1.054,3.593)--(1.234,3.593);
\draw[gp path] (3.930,3.593)--(3.750,3.593);
\node[gp node right] at (0.925,3.593) {$1$};
\gpcolor{color=gp lt color axes}
\gpsetlinetype{gp lt axes}
\gpsetdashtype{gp dt axes}
\gpsetlinewidth{0.50}
\draw[gp path] (1.465,0.432)--(1.465,3.593);
\gpcolor{color=gp lt color border}
\gpsetlinetype{gp lt border}
\gpsetdashtype{gp dt solid}
\gpsetlinewidth{1.00}
\draw[gp path] (1.465,0.432)--(1.465,0.612);
\draw[gp path] (1.465,3.593)--(1.465,3.413);
\node[gp node center] at (1.465,0.216) {N=16};
\gpcolor{color=gp lt color axes}
\gpsetlinetype{gp lt axes}
\gpsetdashtype{gp dt axes}
\gpsetlinewidth{0.50}
\draw[gp path] (2.492,0.432)--(2.492,3.593);
\gpcolor{color=gp lt color border}
\gpsetlinetype{gp lt border}
\gpsetdashtype{gp dt solid}
\gpsetlinewidth{1.00}
\draw[gp path] (2.492,0.432)--(2.492,0.612);
\draw[gp path] (2.492,3.593)--(2.492,3.413);
\node[gp node center] at (2.492,0.216) {N=32};
\gpcolor{color=gp lt color axes}
\gpsetlinetype{gp lt axes}
\gpsetdashtype{gp dt axes}
\gpsetlinewidth{0.50}
\draw[gp path] (3.519,0.432)--(3.519,3.413)--(3.519,3.593);
\gpcolor{color=gp lt color border}
\gpsetlinetype{gp lt border}
\gpsetdashtype{gp dt solid}
\gpsetlinewidth{1.00}
\draw[gp path] (3.519,0.432)--(3.519,0.612);
\draw[gp path] (3.519,3.593)--(3.519,3.413);
\node[gp node center] at (3.519,0.216) {N=64};
\draw[gp path] (1.054,3.593)--(1.054,0.432)--(3.930,0.432)--(3.930,3.593)--cycle;
\node[gp node center,rotate=-270] at (0.204,2.012) {Completion time [sec]};
\gpcolor{rgb color={0.580,0.000,0.827}}
\draw[gp path] (1.465,1.457)--(1.465,2.195);
\draw[gp path] (1.375,1.457)--(1.555,1.457);
\draw[gp path] (1.375,2.195)--(1.555,2.195);
\draw[gp path] (2.492,1.693)--(2.492,2.283);
\draw[gp path] (2.402,1.693)--(2.582,1.693);
\draw[gp path] (2.402,2.283)--(2.582,2.283);
\draw[gp path] (3.519,2.019)--(3.519,2.546);
\draw[gp path] (3.429,2.019)--(3.609,2.019);
\draw[gp path] (3.429,2.546)--(3.609,2.546);
\gpsetpointsize{4.00}
\gppoint{gp mark 1}{(1.465,1.826)}
\gppoint{gp mark 1}{(2.492,1.988)}
\gppoint{gp mark 1}{(3.519,2.282)}
\gpsetpointsize{2.40}
\gppoint{gp mark 7}{(1.465,1.826)}
\gppoint{gp mark 7}{(2.492,1.988)}
\gppoint{gp mark 7}{(3.519,2.282)}
\gpcolor{color=gp lt color border}
\draw[gp path] (1.054,3.593)--(1.054,0.432)--(3.930,0.432)--(3.930,3.593)--cycle;
\gpdefrectangularnode{gp plot 1}{\pgfpoint{1.054cm}{0.432cm}}{\pgfpoint{3.930cm}{3.593cm}}
\end{tikzpicture}
}
					\subfloat[]{\begin{tikzpicture}[gnuplot]
\tikzset{every node/.append style={font={\fontsize{7.0pt}{8.4pt}\selectfont}}}
\path (0.000,0.000) rectangle (4.318,3.810);
\gpcolor{color=gp lt color axes}
\gpsetlinetype{gp lt axes}
\gpsetdashtype{gp dt axes}
\gpsetlinewidth{0.50}
\draw[gp path] (1.054,0.432)--(3.930,0.432);
\gpcolor{color=gp lt color border}
\gpsetlinetype{gp lt border}
\gpsetdashtype{gp dt solid}
\gpsetlinewidth{1.00}
\draw[gp path] (1.054,0.432)--(1.234,0.432);
\draw[gp path] (3.930,0.432)--(3.750,0.432);
\node[gp node right] at (0.925,0.432) {$0.7$};
\gpcolor{color=gp lt color axes}
\gpsetlinetype{gp lt axes}
\gpsetdashtype{gp dt axes}
\gpsetlinewidth{0.50}
\draw[gp path] (1.054,0.959)--(3.930,0.959);
\gpcolor{color=gp lt color border}
\gpsetlinetype{gp lt border}
\gpsetdashtype{gp dt solid}
\gpsetlinewidth{1.00}
\draw[gp path] (1.054,0.959)--(1.234,0.959);
\draw[gp path] (3.930,0.959)--(3.750,0.959);
\node[gp node right] at (0.925,0.959) {$0.75$};
\gpcolor{color=gp lt color axes}
\gpsetlinetype{gp lt axes}
\gpsetdashtype{gp dt axes}
\gpsetlinewidth{0.50}
\draw[gp path] (1.054,1.486)--(3.930,1.486);
\gpcolor{color=gp lt color border}
\gpsetlinetype{gp lt border}
\gpsetdashtype{gp dt solid}
\gpsetlinewidth{1.00}
\draw[gp path] (1.054,1.486)--(1.234,1.486);
\draw[gp path] (3.930,1.486)--(3.750,1.486);
\node[gp node right] at (0.925,1.486) {$0.8$};
\gpcolor{color=gp lt color axes}
\gpsetlinetype{gp lt axes}
\gpsetdashtype{gp dt axes}
\gpsetlinewidth{0.50}
\draw[gp path] (1.054,2.013)--(3.930,2.013);
\gpcolor{color=gp lt color border}
\gpsetlinetype{gp lt border}
\gpsetdashtype{gp dt solid}
\gpsetlinewidth{1.00}
\draw[gp path] (1.054,2.013)--(1.234,2.013);
\draw[gp path] (3.930,2.013)--(3.750,2.013);
\node[gp node right] at (0.925,2.013) {$0.85$};
\gpcolor{color=gp lt color axes}
\gpsetlinetype{gp lt axes}
\gpsetdashtype{gp dt axes}
\gpsetlinewidth{0.50}
\draw[gp path] (1.054,2.539)--(3.930,2.539);
\gpcolor{color=gp lt color border}
\gpsetlinetype{gp lt border}
\gpsetdashtype{gp dt solid}
\gpsetlinewidth{1.00}
\draw[gp path] (1.054,2.539)--(1.234,2.539);
\draw[gp path] (3.930,2.539)--(3.750,2.539);
\node[gp node right] at (0.925,2.539) {$0.9$};
\gpcolor{color=gp lt color axes}
\gpsetlinetype{gp lt axes}
\gpsetdashtype{gp dt axes}
\gpsetlinewidth{0.50}
\draw[gp path] (1.054,3.066)--(3.930,3.066);
\gpcolor{color=gp lt color border}
\gpsetlinetype{gp lt border}
\gpsetdashtype{gp dt solid}
\gpsetlinewidth{1.00}
\draw[gp path] (1.054,3.066)--(1.234,3.066);
\draw[gp path] (3.930,3.066)--(3.750,3.066);
\node[gp node right] at (0.925,3.066) {$0.95$};
\gpcolor{color=gp lt color axes}
\gpsetlinetype{gp lt axes}
\gpsetdashtype{gp dt axes}
\gpsetlinewidth{0.50}
\draw[gp path] (1.054,3.593)--(3.930,3.593);
\gpcolor{color=gp lt color border}
\gpsetlinetype{gp lt border}
\gpsetdashtype{gp dt solid}
\gpsetlinewidth{1.00}
\draw[gp path] (1.054,3.593)--(1.234,3.593);
\draw[gp path] (3.930,3.593)--(3.750,3.593);
\node[gp node right] at (0.925,3.593) {$1$};
\gpcolor{color=gp lt color axes}
\gpsetlinetype{gp lt axes}
\gpsetdashtype{gp dt axes}
\gpsetlinewidth{0.50}
\draw[gp path] (1.465,0.432)--(1.465,3.593);
\gpcolor{color=gp lt color border}
\gpsetlinetype{gp lt border}
\gpsetdashtype{gp dt solid}
\gpsetlinewidth{1.00}
\draw[gp path] (1.465,0.432)--(1.465,0.612);
\draw[gp path] (1.465,3.593)--(1.465,3.413);
\node[gp node center] at (1.465,0.216) {N=16};
\gpcolor{color=gp lt color axes}
\gpsetlinetype{gp lt axes}
\gpsetdashtype{gp dt axes}
\gpsetlinewidth{0.50}
\draw[gp path] (2.492,0.432)--(2.492,3.593);
\gpcolor{color=gp lt color border}
\gpsetlinetype{gp lt border}
\gpsetdashtype{gp dt solid}
\gpsetlinewidth{1.00}
\draw[gp path] (2.492,0.432)--(2.492,0.612);
\draw[gp path] (2.492,3.593)--(2.492,3.413);
\node[gp node center] at (2.492,0.216) {N=32};
\gpcolor{color=gp lt color axes}
\gpsetlinetype{gp lt axes}
\gpsetdashtype{gp dt axes}
\gpsetlinewidth{0.50}
\draw[gp path] (3.519,0.432)--(3.519,3.413)--(3.519,3.593);
\gpcolor{color=gp lt color border}
\gpsetlinetype{gp lt border}
\gpsetdashtype{gp dt solid}
\gpsetlinewidth{1.00}
\draw[gp path] (3.519,0.432)--(3.519,0.612);
\draw[gp path] (3.519,3.593)--(3.519,3.413);
\node[gp node center] at (3.519,0.216) {N=64};
\draw[gp path] (1.054,3.593)--(1.054,0.432)--(3.930,0.432)--(3.930,3.593)--cycle;
\node[gp node center,rotate=-270] at (0.204,2.012) {Completion time [sec]};
\gpcolor{rgb color={0.580,0.000,0.827}}
\draw[gp path] (1.465,1.375)--(1.465,2.197);
\draw[gp path] (1.375,1.375)--(1.555,1.375);
\draw[gp path] (1.375,2.197)--(1.555,2.197);
\draw[gp path] (2.492,1.719)--(2.492,2.393);
\draw[gp path] (2.402,1.719)--(2.582,1.719);
\draw[gp path] (2.402,2.393)--(2.582,2.393);
\draw[gp path] (3.519,1.994)--(3.519,2.689);
\draw[gp path] (3.429,1.994)--(3.609,1.994);
\draw[gp path] (3.429,2.689)--(3.609,2.689);
\gpsetpointsize{4.00}
\gppoint{gp mark 1}{(1.465,1.786)}
\gppoint{gp mark 1}{(2.492,2.056)}
\gppoint{gp mark 1}{(3.519,2.341)}
\gpsetpointsize{2.40}
\gppoint{gp mark 7}{(1.465,1.786)}
\gppoint{gp mark 7}{(2.492,2.056)}
\gppoint{gp mark 7}{(3.519,2.341)}
\gpcolor{color=gp lt color border}
\draw[gp path] (1.054,3.593)--(1.054,0.432)--(3.930,0.432)--(3.930,3.593)--cycle;
\gpdefrectangularnode{gp plot 1}{\pgfpoint{1.054cm}{0.432cm}}{\pgfpoint{3.930cm}{3.593cm}}
\end{tikzpicture}
}
			\end{tabular}}
		\end{tabular}
		\caption{Completion time of fiber-path (a) establishment and (b) release on emulation environment.}
		\label{fig:results-emulation}
	\end{figure}

	\subsection{Evaluation on scalability in large-scale emulated networks}
	We also evaluated the scalability of our controller in emulated large-scale networks. Since the construction of actual large-scale OCS-based networks was difficult, we implemented and prepared the emulated OCSes and their drivers on the controller. We set up the emulated network as shown in Fig.~\ref{fig:exp-emu}; terminals A and Z are attached to the edge of the networks, and there are three routes consisting of $N$ OCSes between them. The configuration time of each OCS was randomly generated from a normal distribution with a mean of $0.7$ seconds and standard deviation of $0.07$ in accordance with actual configuration time for path operation investigated in Section~\ref{sec:experimental-validation}.\ref{subsec:exp-nbi}.
	
	We then evaluated the completion time required for fiber-path establishment and release by setting $N$ to $16$, $32$, and $64$. In each case, there were $44$, $92$, and $188$ OCS nodes in the network. We executed these operations $10$ times in each route, and the averaged values are shown in Fig.~\ref{fig:results-emulation}. As shown in the figure, the completion time become longer as $N$ grew in each fiber-path operation. Considering that the completion time depends on the maximum configuration time among the OCSes on the path due to parallel configurations, it could be likely to take larger values as $N$ increases. However, we found that the completion time was always within $1.0$ second under all settings. In addition, we found that our controller could manage the large-scale networks with many OCS nodes. Though further increase of OCS nodes might lead to longer configuration delays, more than $64$ OCSes on a path would be rare case. From these results, we could confirm the high scalability of our controller and the effectiveness of our designed framework, especially the parallel and atomic configurations of OCSes.

	\subsection{Demonstration of managing metro-access converged networks with OCSes}
\begin{figure}[t]
		\centering
		\includegraphics[width=.48\textwidth]{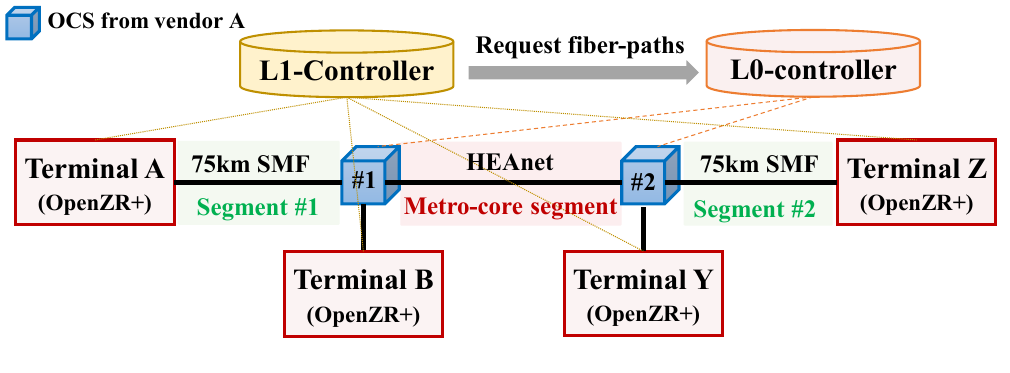}
		\caption{Experimental testbed of metro-access converged networks with OCSes.}
		\label{fig:exp-amc}
	\end{figure}

	We finally demonstrate the feasibility of our controller for managing metro-access converged networks with OCSes in a field environment. Fig.~\ref{fig:exp-amc} shows the experimental setup in the CONNECT Centre, a city-scale testbed in Dublin, Ireland. We constructed two access segments (segments \#1 and \#2) using 75km spool fibers in the CONNECT Centre, and connected them to the HEAnet (Ireland’s National Research and Education Network) live production Dublin metro network for the metro-core segment. We then attached the open optical equipment AGR400 from Edgecore~\cite{agr400} with OpenZR+ compatible coherent pluggable modules as terminals. The open-source network operating system (NOS) called Goldstone, which was developed in Telecom Infra Project Open Optical Packet \& Transport (TIP-OOPT)~\cite{anazawa2024latest, goldstone}, was installed on the AGR400 and used for controlling the terminals. The terminals were configured by our developed L1-(Layer-1) controller. Using this field environment, we demonstrate that our L0-controller efficiently sets up fiber paths between source and destination terminals upon receiving the requests from the L1-controller. In this case, the L0-controller acts as a server for providing fiber paths.
	
	The procedure of this experiment is as follows. First, the ROADMs and amplifiers on the line systems were set up beforehand to accommodate signals from the terminals. Second, we set up fiber paths between terminals A and B and terminals Z and Y, upon receiving a path setup request from L1-controller. Then, the link-by-link probing for estimating the QoT of the access segments~\cite{nishizawa2024fast, kaeval2024operation} was performed by our L1-controller. After that, we set up a fiber path between terminals B and Y by L0-controller for estimating the QoT of a metro-core segment. Finally, we set up a fiber path between terminals A and Z using L0-controller and set up the wavelength path by L1-controllerby tuning transmission mode, wavelength, and transmission power.
	
	Through this experiment, we found that the fiber paths were always set up within $1.0$ second, which demonstrates the efficient fiber-path operation by our L0-controller. It also demonstrates the feasibility and validity of our controller for managing metro-access converged networks with OCSes.

	\subsection{Future directions}
	Through designing and implementing the MV-OCS controller in this work, we identified three important future directions.
	
	First, though we evaluated our MV-OCS controller in terms of its control functionality and performance, the power dynamics due to optical switching and restoration (e.g. as shown in \cite{mo2017dual}) were not evaluated and might lead to additional configuration delays due to tuning of the physical layer elements in problematic scenarios. Impacts were not observed for the cases considered here. Further studies on the power dynamics issues would be an interesting topic when operating OCS-based networks by the open MV-OCS controller.
	
	Second, the unified models for alarms notified from OCSes themselves due to equipment failures should be considered. Since current OCS products have no unified SBI for handling equipment alarms, implementing and evaluating such a unified SBI could be an important future direction.
	
	Finally, further discussions and research effort on fault tolerant network management using the SDN controller would be required. Though we proposed a series of basic sanity check and failure recovery mechanisms on our MV-OCS controller, there might be problematic cases that still cannot be addressed by our proposed framework. An example of such a case includes no alarm from OCSes or terminals because of their silent failures. In this case, more sophisticated failure localization or recovery mechanisms could be made possible by leveraging the layer 2 (or a higher networking layer) controller. Realizing such a fault tolerant mechanisms should be a very important research topic.

	\section{Concluding Remarks}
	\label{sec:concluding-remarks}
	In this paper, we studied a software-defined networking (SDN) controller for open optical-circuit-switched networks. We first discussed the OCS use cases and operation tasks on the OCS-based networks to clarify the controller requirements. We then studied a multi-vendor (MV) OCS controller framework including north-bound interface (NBI), south-bound interface (SBI) for operating MV-OCSes, and the internal functions and data models inside the controller, to satisfy the derived controller requirements. Experimental results on a physical testbed constructed by actual MV-OCSes revealed the feasibility and validity of our developed controller; it satisfied all the controller requirements and quickly and safely controlled fiber paths on OCS-based networks. In addition, we successfully operated optical-circuit-switched metro-access converged networks using our controller in live production metro networks. The fiber paths were always established and released within $1.0$ second. Our future work includes promoting our MV-OCS controller framework through the collaboration with OCS vendors and standard-defining organizations (SDOs). The collaboration with traditional SDN controllers would also be an interesting topic for further acceleration in the research fields related to open optical networking.

	\section{Acknowledgments}
	Portions of this work were presented at the IEEE Opto-Electronics and Communications Conference (OECC) in 2024.
	
	These research results were obtained from the grant program (No. JPJ012368G50201) by National Institute of Information and Communications Technology (NICT), Japan.
	\bibliography{./References/ref}
	
	\bigskip
	
	\section{Author Biographies}
	
	\setlength\intextsep{0pt}
	
	\medskip
	
	\noindent \textbf{Kazuya Anazawa} is a researcher at NTT Network Innovation Laboratories, Japan. He received B.E. and M.E. degrees in computer science and engineering from the University of Aizu, Japan in 2016 and 2018. In 2018, he joined NTT Network Innovation Laboratories.
	His research interests include optical network design and autonomous control of optical networks.
	
	\medskip
	\medskip
	
	\noindent \textbf{Takeru Inoue} is a distinguished researcher at NTT Network Innovation Laboratories, Japan. He received B.E. (1998) and M.E. (2000) degrees in engineering science and Ph.D. degree in information science (2006) from Kyoto University, Japan. In 2000, he joined NTT Laboratories. From 2011 to 2013, he was an ERATO researcher with the Japan Science and Technology Agency, where his research focused on algorithms and data structures. His research interests widely cover algorithmic approaches in communication networks. He has received several prestigious awards, including the Best Paper Award of the Asia-Pacific Conference on Communications in 2005, the Best Paper Award of the IEEE International Conference on Communications in 2016, the Best Paper Award of the IEEE Global Communications Conference in 2017, the Best Paper Award of IEEE Reliability Society Japan Joint Chapter in 2020, the IEEE Asia/Pacific Board Outstanding Paper Award in 2020, and the IEICE Paper of the Year in 2021. He serves as an associate editor of the IEEE Transactions on Network and Service Management.
	
	\medskip
	\medskip
	
	\noindent \textbf{Toru Mano} received B.E. and M.E. degrees from the University of Tokyo, Japan in 2009 and 2011 and Ph.D. degree in computer science and information technology from Hokkaido University, Japan in 2020. He joined NTT Network Innovation Laboratories, Japan, in 2011, where he is a senior research engineer. His research interests are network architectures, network optimization, and softwarization of networking.
	He was a recipient of the IEICE Paper of the Year in 2021 and the Best Paper Award at European Conference on Optical Communications 2023.
	He is a member of IEICE and Operations Research Society of Japan.
	
	\medskip
	\medskip
	
	\noindent \textbf{Hiroshi Ou} received B.E. and M.E. degrees in computer science and engineering from Waseda University, Tokyo, Japan, in 2009 and 2011. He joined NTT Access Network Service Systems Laboratories, Yokosuka, Japan in 2009, where he researched optical access networks and systems and optical-wireless converged networks. From 2016 to 2020, he was with NTT DOCOMO, where he developed LTE and 5G base stations. In 2020, he returned to NTT Access Network Service Systems Laboratories. He is a member of IEICE.
	
	\medskip
	\medskip
	
	\noindent \textbf{Hirotaka Ujikawa} is a senior research engineer in the Optical Access System Project at NTT Access Network Service Systems Laboratories. He received a B.E. and M.E. in computer science from Waseda University, Tokyo, Japan in 2007 and 2009, and a Ph.D. in information science from Tohoku University, Miyagi, Japan in 2017. He joined NTT in 2009, where he has been researching and developing optical access systems. His current research interests include dynamic bandwidth allocation for low latency services and dynamic sleep scheduling for energy efficient access systems. He is a member of IEICE.
	
	\medskip
	\medskip
	
	\noindent \textbf{Dmitrii Briantcev} is a CONNECT Research Assistant working at Trinity College Dublin, Ireland. He obtained his bachelor degree in radiophysics from Saint Petersburg State University (SPbSU), Russia in 2018. He then joined the Communication Theory Lab (CTL) at King Abdullah University of Science and Technology (KAUST), Saudi Arabia obtaining MS and PhD degrees in the field of electrical and computer engineering, focusing on physical simulations and machine learning applications for Free Space Optical (FSO) communication systems. In June 2023, Dmitrii joined CONNECT to work on quantum optical communications under the IrelandQCI project.
	
	\medskip
	\medskip
	
	\noindent \textbf{Sumaiya Binte Ali} is a CONNECT Researcher at Trinity College Dublin, Ireland where she is currently pursuing her Ph.D. Her research focuses on cross-layer control, multi-domain optical networks, and programmable switches designed for high-capacity applications and emerging technologies such as 6G. Her ultimate goal is to facilitate the connection between academic research and the practical deployment of optical communication solutions in community-driven smart cities.
	
	\medskip
	\medskip
	
	\noindent \textbf{Devika Dass} is a CONNECT Research Fellow at Trinity College Dublin, Ireland. Her research interests include advanced modulation formats and their applications in converged optical access networks, characterization of Si-based photonic devices, photonic mmWave generation, short-reach intra-DC connects, and passive optical networks. She received her Bachelor of Technology in electronics and telecommunication from Amity University, India in 2014. In 2015, she received her Master’s in communication engineering from VIT University, Vellore, India, during which she interned at CSIR-National Physical Laboratory. She has modeled the attenuation of free space optical communication (FSOC) in urban environments and experimentally analyzed the attenuation of FSOC in the presence of fog. Later in 2017, she joined IITD as a research fellow and analyzed radio-over-fiber technology for different modulation schemes in simulations and experiments. She completed her Ph.D. in the Radio and Optical Communications Lab at Dublin City University, Ireland in 2023, and her research presented the deployment of a novel, integrable, ultra-flexible and low noise SiP optical source to facilitate reconfigurable millimeter wave frequency transmission systems, multi-service environment by pairing it with optical switch fabric and high bandwidth intra-datacenter interconnection.
	
	\medskip
	\medskip
	
	\noindent \textbf{Eoin Kenny} is the innovation, research and development manager at HEAnet (Ireland’s National Research \& Education Network). His research interests are in the areas of advanced network architectures, quantum communications, distribution of timing services, fiber  sensing and control, and management planes for future networks. He received a degree in electronic engineering from University College Dublin, Ireland in 1993.
	
	\medskip
	\medskip
	
	\noindent \textbf{Hideki Nihsizawa} is a senior research engineer, supervisor, at NTT Network Innovation Laboratories, Japan. He received B.E. and M.E. degrees in physics from Chiba University, Chiba, Japan, in 1994 and 1996. In 1996, he joined NTT Laboratories, Japan, where he has been engaged in research on open and disaggregated optical systems.
	He serves as a technical co-lead of the open optical \& packet transport disaggregated optical system (OOPT-DOS) group of the Telecom Infra Project, and a coordinator at the IOWN Global Forum Open APN Functional Architecture task force.
	
	\medskip
	\medskip
	
	\noindent \textbf{Yoshiaki Sone} is a senior research engineer, supervisor, at NTT Network Innovation Laboratories, Japan. He received an M.E. in electronics engineering from Tohoku University, Miyagi, Japan in 2003. Since joining NTT the same year, he has been involved in research and technological development of optical transport systems and associated network engineering and standardization. His primary focus is on the development of open optical transport systems as well as related standardization activities including the OpenROADM MSA.
	
	\medskip
	\medskip
	
	\noindent \textbf{Marco Ruffini} received his M.Eng. in telecommunications engineering in 2002 from Polytechnic University of Marche, Italy. After working as a research scientist for Philips in Germany, he joined Trinity College Dublin (TCD), Ireland in 2005, where he received his Ph.D. in 2007 and is currently a full professor. He is a principal investigator (PI) of both the CONNECT Telecommunications Research Center at TCD and the IPIC Photonics Integration Centre headquartered at the National Tyndall Institute. Prof. Ruffini is currently involved in several Science Foundation Ireland (SFI) and H2020 projects. He leads the Optical and Radio Network Architecture Group at TCD and the OpenIreland Beyond 5G Testbed Research Infrastructure. His main research is in the area of 5G optical networks, where he carries out pioneering work on the convergence of fixed-mobile and access-metro networks, and on the virtualization of next-generation networks. He has been invited to share his vision through several keynote addresses and talks at major international conferences across the world. He has recently started working also on quantum networking, where he is collaborating with the US Center for Quantum Networks (CQN). He has authored over 200 international publications, holds 10 patents, contributed to industry standards, secured research funding for over €14 million, and contributed the novel virtual Dynamic Bandwidth Allocation (vDBA) concept to the Broadband Forum standardization body.
	
	\medskip
	\medskip
	
	\noindent \textbf{Daniel Kilper} is a director of the CONNECT Centre at Trinity College Dublin, Ireland and principal investigator (PI). He is also a professor of future communication networks in the School of Engineering in Trinity College Dublin. Prof. Kilper received his PhD in physics from the University of Michigan, USA in 1996. From 2000 to 2013, he was a member of the technical staff at Bell Labs. He is a senior member of IEEE, a topical area editor for the IEEE Transactions on Green Communications and Networking (TGCN), and chairs the optics working group in the IEEE International Network Generations Roadmap. He was recognized as a 2019 NIST Communications Technology Lab Innovator, holds 11 issued patents, and has authored 6 book chapters and more than 167 peer-reviewed publications. His research is aimed at solving fundamental and real-world problems in communication networks to create a faster, more affordable, and energy efficient Internet, addressing interdisciplinary challenges for smart cities, sustainability, and digital equity. Prof. Kilper comes to CONNECT from the University of Arizona, USA where he held a research professorship in the College of Optical Sciences, and a joint appointment in electrical and computer engineering. He also holds an adjunct faculty position in the Data Science Institute at Columbia University, USA where he is a co-PI on the COSMOS advanced wireless testbed. He was lead PI on the founding project of SFI’s US-Ireland Research and Development Partnership and has served in leadership positions in multiple international university-industry research centers including participating in CONNECT’s precursor, CTVR, while he was at Bell Labs.
	
	\medskip
	\medskip
	
	\noindent \textbf{Eiji Oki} received B.E. and M.E. degrees in instrumentation engineering and Ph.D. degree in electrical engineering from Keio University, Yokohama, Japan, in 1991, 1993, and 1999. From 1993 to 2008, he was a researcher at NTT Laboratories, Tokyo, Japan. From 2000 to 2001, he was a visiting scholar at Polytechnic University, Brooklyn, NY, USA. He was a professor at the University of Electro-Communications, Tokyo, from 2008 to 2017. Since 2017, he has been a professor at Kyoto University, Kyoto, Japan. He has been active in the standardization of the path computation element in Internet Engineering Task Force (IETF). He has authored/contributed to 12 IETF RFCs. His research interests include routing, switching, protocols, optimization, and traffic engineering in communication and information networks. He is a fellow of IEEE and IEICE.
	
	\medskip
	\medskip
	
	\noindent \textbf{Koichi Takasugi} is an executive research engineer, director, and head of the Frontier Communication Laboratory, NTT Network Innovation Laboratories, Japan. He received a B.E. in computer science from Tokyo Institute of Technology, M.E. from JAIST, and Ph.D. in engineering from Waseda University, Tokyo, Japan in 1995, 1997, and 2004. He was involved in the design and standardization of the next-generation network architecture. He has implemented and installed super high-density Wi-Fi systems in several soccer stadiums. He was also active in the artificial intelligence field, such as diagnosing diabetes by machine learning. He is currently leading research on the network architecture and protocols in optical and wireless transport networks.
	
\end{document}